\documentclass[aps,showpacs,twocolumn,superscriptaddress]{revtex4} 
 
\usepackage{graphicx} 
\usepackage{dcolumn} 
\usepackage{amsmath} 
 
\newcommand{\sfrac}[2]{\mbox{\footnotesize $\displaystyle \frac{#1}{#2}$}} 
 
\begin{document} 

 
\preprint{ANL-PHY-9968-TH-2001} 
 
\title{Nucleon mass and pion loops} 
 
 
\author{M.B.~Hecht} 
\affiliation{Physics Division, Bldg 203, Argonne National Laboratory, Argonne 
Illinois 60439-4843\vspace*{1ex}} 
%
\author{M.~Oettel} 
\affiliation{Special Research Centre for the Subatomic Structure of Matter and 
Department of Physics and Mathematical Physics, University of Adelaide, 
Adelaide SA 5005, Australia\vspace*{1ex}} 
\author{C.D.~Roberts} 
\affiliation{Physics Division, Bldg 203, Argonne National Laboratory, Argonne 
Illinois 60439-4843\vspace*{1ex}} 
\author{S.M.~Schmidt} 
\affiliation{Institut f\"ur Theoretische Physik, Universit\"at T\"ubingen, Auf 
der Morgenstelle 14, D-72076 T\"ubingen, Germany\vspace*{1ex}} 
\author{P.C.~Tandy} 
\affiliation{Center for Nuclear Research, Department of Physics, Kent State 
University, Kent OH 44242\vspace*{1ex}} 
\author{A.W.~Thomas\vspace*{1ex}} 
\affiliation{Special Research Centre for the Subatomic Structure of Matter and 
Department of Physics and Mathematical Physics, University of Adelaide, 
Adelaide SA 5005, Australia\vspace*{1ex}} 
%
 
 
\begin{abstract} 
\rule{0ex}{3ex} 
Poincar\'e covariant Faddeev equations for the nucleon and $\Delta$ are
solved to illustrate that an internally consistent description in terms of
confined-quark and nonpointlike confined-diquark-correlations can be
obtained.  $\pi N$-loop induced self-energy corrections to the nucleon's mass
are analysed and shown to be independent of whether a pseudoscalar or
pseudovector coupling is used.  Phenomenological constraints suggest that
this self-energy correction reduces the nucleon's mass by up to several
hundred MeV. That effect does not qualitatively alter the picture, suggested
by the Faddeev equation, that baryons are quark-diquark composites. However,
neglecting the $\pi$-loops leads to a quantitative overestimate of the
nucleon's axial-vector diquark component.
\end{abstract} 
 
\maketitle 
 
\section{Introduction} 
\label{introduction} 
Contemporary experimental facilities employ large momentum transfer reactions
to probe the structure of hadrons and thereby attempt to elucidate the role
played by quarks and gluons in building them.  Since the proton is a readily
accessible target its properties have been studied most
extensively~\cite{Thomas:kw}. Hence an understanding of a large fraction of
the available data requires a Poincar\'e covariant theoretical description of
the nucleon.
 
At its simplest the nucleon is a nonperturbative three-body bound-state 
problem, an exact solution of which is difficult to obtain even if the 
interactions are known.  Hitherto, therefore, phenomenological mean-field 
models have been widely employed to describe nucleon structure; e.g., soliton 
models~\cite{wilets,Birse:cx,reinhardTS} and constituent-quark 
models~\cite{tonyCBM,tonyANU,simon}.  These models are most naturally applied 
to processes involving small momentum transfer ($q^2< M^2$, $M$ is the nucleon 
mass) and, as commonly formulated, their applicability may be extended to 
processes involving larger momentum transfer by working in the Breit 
frame~\cite{gerrytony}. Alternatively, one could define an equivalent, Galilean 
invariant Hamiltonian and reinterpret that as the Poincar\'e invariant mass 
operator for a quantum mechanical theory~\cite{fritz} but this path is less 
well travelled. 
 
Another approach is to describe the nucleon via a Poincar\'e covariant Faddeev 
equation.  That, too, requires an assumption about the interaction between 
quarks.  An analysis~\cite{regbos} of the Global Colour 
Model~\cite{gcm,petergcm,gunnergcm} suggests that the nucleon can be viewed as 
a quark-diquark composite.  Pursuing that picture yields~\cite{regfe} a 
Faddeev equation, in which two quarks are always correlated as a 
colour-antitriplet diquark quasiparticle (because ladder-like gluon exchange is 
attractive in the $\bar 3_c$ quark-quark scattering channel) and binding in the 
nucleon is effected by the iterated exchange of roles between the dormant and 
diquark-participant quarks. 
 
A first numerical study of this Faddeev equation for the nucleon was reported 
in Ref.~\cite{cjbfe}, and following that there have been numerous more 
extensive analyses; e.g., Refs.~\cite{bentz,oettel}.  In particular, the 
formulation of Ref.~\cite{oettel} employs confined quarks, and confined, 
pointlike-scalar and -axial-vector diquark correlations, to obtain a spectrum 
of octet and decuplet baryons in which the rms-deviation between the calculated 
mass and experiment is only $2$\%.  The model also reproduces nucleon form 
factors over a large range of momentum transfer~\cite{oettel2}, and its 
descriptive success in that application is typical of such Poincar\'e covariant 
treatments; e.g., Refs.~\cite{jacquesA,jacquesmyriad,cdrqciv,nedm}. 
 
However, these successes might themselves indicate a flaw in the application of 
the Faddeev equation to the nucleon.  For example, in the context of 
spectroscopy, studies using the Cloudy Bag Model (CBM)~ \cite{tonyCBM} indicate 
that the dressed-nucleon's mass receives a negative contribution of as much as 
$300$-$400\,$MeV from pion self-energy corrections; i.e., $\delta M_+ = -300\,$ 
to $-400\,$MeV~\cite{tonyANU,bruceCBM}. Furthermore, a perturbative study, 
using the Faddeev equation, of the mass shift induced by pointlike-$\pi$ 
exchange between the quark and diquark constituents of the nucleon obtains 
$\delta M_+ = -150$ to $-300\,$MeV~\cite{ishii}. Unameliorated these mutually 
consistent results would much diminish the value of the $2$\% spectroscopic 
accuracy obtained using only quark and diquark degrees of freedom. 
 
It is thus apparent that the size and qualitative impact of the pionic
contribution to the nucleon's mass may provide material constraints on the
development of a realistic quark-diquark picture of the nucleon, and its
interpretation and application.  Our article is an exploration of this
possibility and we aim to clarify the model dependent aspects. We emphasise,
in addition, that chiral corrections to baryon magnetic moments and charge
radii are also important~\cite{radiiCh}, and their model-independent features
furnish additional constraints on any quark model, including those based on
the Faddeev equation, thereby guiding their improvement.  We note, too, that
lattice-QCD studies of baryon masses, especially as a function of the
current-quark mass~\cite{LQCD}, also provide information that can guide these
considerations; e.g., a recent lattice-QCD exploration of the connection
between $N$ and $\Delta$ masses is consistent with the pion self-energies
described above~\cite{Young:2001nc}.
 
In Sec.~\ref{sec:Fad} we recapitulate on the Faddeev equation and its
solution for the $N$ and $\Delta$ in a simple model.
Section~\ref{sec:piloop} discusses model-independent aspects of the
Dyson-Schwinger equation (DSE)~\cite{cdragw} that describes the pionic
correction to the $N$'s self-energy and therein we also present exemplary
estimates for the magnitude of the effect.  Section~\ref{sec:epil} is an
epilogue.
 
\section{Faddeev Equation} 
\label{sec:Fad} 
The properties of light pseudoscalar and vector mesons are well described by
a renormalisation-group-improved rainbow-ladder truncation of QCD's
DSEs~\cite{mr97,pieterrho,pieterGamma}, and the study of baryons via the
solution of a Poincar\'e covariant Faddeev equation is a desirable extension
of the approach.  The derivation of a Faddeev equation for the bound state
contribution to the three quark scattering kernel is possible because the
same kernel that describes mesons so well is also strongly attractive for
quark-quark scattering in the colour-antitriplet channel (see
Sec.~\ref{subsubsec:Gamma}).  And it is a simple consequence of the
Clebsch-Gordon series for quarks in the fundamental representation of
$SU_c(3)$:
\begin{equation} 
3_c \otimes 3_c \otimes 3_c = (\bar 3_c \oplus 6_c) \otimes 3_c = 1_c \oplus 
8_c^\prime \oplus 8_c \oplus 10_c\,, 
\end{equation} 
that any two quarks in a colour singlet bound state must constitute a relative 
colour antitriplet.  This supports a truncation of the three-body problem 
wherein the interactions between two selected quarks are added to yield a 
quark-quark scattering matrix, which is then approximated as a sum over all 
possible diquark pseudoparticle terms: Dirac-scalar $+$ -pseudovector $+ 
[\ldots]$ -- essentially a separable two-body interaction~\cite{Modern}.  A 
Faddeev equation follows, which describes the three-body bound-state as a 
composite of a dressed-quark and nonpointlike diquark with an iterated exchange 
of roles between the dormant and diquark-participant quarks. The bound-state is 
represented by a Faddeev amplitude: 
\begin{equation} 
\Psi = \Psi_1 + \Psi_2 + \Psi_3 \,, 
\end{equation} 
where the subscript identifies the dormant quark and, e.g., $\Psi_{1,2}$ are 
obtained from $\Psi_3$ by a correlated, cyclic permutation of all the quark 
labels. 
 
The Faddeev equation is simplified further by retaining only the lightest 
diquark correlations in the representation of the quark-quark scattering 
matrix.  A simple, Goldstone-theorem-preserving, rainbow-ladder 
DSE-model~\cite{conradsep} yields the following diquark pseudoparticle masses 
(isospin symmetry is assumed): 
\begin{equation} 
\label{dqmass} 
\begin{array}{l|cccc} 
(qq)_{J^P}           & (ud)_{0^+} & (us)_{0^+}  & (uu)_{1^+}& (us)_{1^+}\\ 
 m_{qq}\,({\rm GeV}) & 0.74       & 0.88        & 0.95      & 1.05 \\\hline 
(qq)_{J^P}           & (ss)_{1^+} & (uu)_{1^-} & (us)_{1^-} & (ss)_{1^-}  \\ 
 m_{qq}\,({\rm GeV}) & 1.13 & 1.47 & 1.53 & 1.64 
\end{array} 
\end{equation} 
The mass ordering is characteristic and model-independent (cf.\ 
Refs.~\cite{gunner,pieterdq}, lattice-QCD estimates~\cite{latticediquark} and 
studies of the spin-flavour dependence of parton 
distributions~\cite{Close:br}), and indicates that a study of the $N$ and 
$\Delta$ must retain at least the scalar and pseudovector $(uu)$- and 
$(ud)$-correlations if it is to be accurate.  (Of course, the spin-$3/2$ 
$\Delta$ is inaccessible unless pseudovector correlations are retained.) 
 
\subsection{Model for the Nucleon} 
\label{subsec:nucleon} 
To provide a concrete illustration and make our presentation self-contained we 
consider a simple model~\cite{cdrqciv} wherein the nucleon is a sum of scalar 
and pseudovector diquark correlations: 
\begin{equation} 
\label{Psi} \Psi_3(p_i,\alpha_i,\tau_i) = \Psi_3^{0^+} + \Psi_3^{1^+}, 
\end{equation} 
with $(p_i,\alpha_i,\tau_i)$ the momentum, spin and isospin labels of the 
quarks constituting the nucleon, and $P=p_1+p_2+p_3$ the nucleon's total 
momentum.  The scalar diquark component in Eq.~(\ref{Psi}) is 
\begin{eqnarray} 
\nonumber 
\lefteqn{\Psi_3^{0^+}(p_i,\alpha_i,\tau_i)=}\\ 
\nonumber && [\Gamma^{0^+}(\frac{1}{2}p_{[12]};K)]_{\alpha_1 
\alpha_2}^{\tau_1 \tau_2}\, \Delta^{0^+}(K) \,[{\cal S}(\ell;P) u(P)]_{\alpha_3}^{\tau_3}\,,\\ 
\label{calS} && 
\end{eqnarray} 
where~\cite{fn:Euclidean}: the spinor satisfies 
\begin{equation}
(i\gamma\cdot P + M)\, u(P) =0= \bar u(P)\, (i\gamma\cdot P + M)\,,
\end{equation}
with $M$ the mass obtained in solving the Faddeev equation, and is also a
spinor in isospin space with $\varphi_+= {\rm col}(1,0)$ for the proton and
$\varphi_-= {\rm col}(0,1)$ for the neutron; $K= p_1+p_2=: p_{\{12\}}$,
$p_{[12]}= p_1 - p_2$, $\ell := (-p_{\{12\}} + 2 p_3)/3$; $\Delta^{0^+}(K)$
is a pseudoparticle propagator for the scalar diquark formed from quarks $1$
and $2$, and $\Gamma^{0^+}\!$ is a Bethe-Salpeter-like amplitude describing
their relative momentum correlation; and ${\cal S}$, a $4\times 4$ Dirac
matrix, describes the relative quark-diquark momentum correlation.  (${\cal
S}$, $\Gamma^{0^+}$ and $\Delta^{0^+}$ are discussed below.)  The
pseudovector component is
\begin{eqnarray} 
\nonumber 
\lefteqn{ \Psi^{1^+}(p_i,\alpha_i,\tau_i)= }\\ 
\nonumber &&  [{\tt t}^i\,\Gamma_\mu^{1^+}(\frac{1}{2}p_{[12]};K)]_{\alpha_1 
\alpha_2}^{\tau_1 \tau_2}\,\Delta_{\mu\nu}^{1^+}(K)\, 
[{\cal A}^{i}_\nu(\ell;P) u(P)]_{\alpha_3}^{\tau_3}\,,\\ 
\label{calA} && 
\end{eqnarray} 
where the symmetric isospin-triplet matrices are 
\begin{equation} 
{\tt t}^+ = \frac{1}{\surd 2}(\tau^0+\tau^3) \,,\; 
{\tt t}^0 = \tau^1\,,\; 
{\tt t}^- = \frac{1}{\surd 2}(\tau^0-\tau^3)\,, 
\end{equation} 
with $(\tau^0)_{ij}=\delta_{ij}$ and $\tau^{1,3}$ the usual Pauli matrices, 
and the other elements in Eq.~(\ref{calA}) are obvious generalisations of 
those in Eq.~(\ref{calS}). 
 
The colour antisymmetry of $\Psi_3$ is implicit in $\Gamma^{J^P}\!$, with the 
Levi-Civita tensor, $\epsilon_{c_1 c_2 c_3}$, expressed via the antisymmetric 
Gell-Mann matrices; i.e., defining 
\begin{equation} 
\{H^1=i\lambda^7,H^2=-i\lambda^5,H^3=i\lambda^2\}\,, 
\end{equation} 
then $\epsilon_{c_1 c_2 c_3}= (H^{c_3})_{c_1 c_2}$.  (See 
Eqs.~(\ref{Gamma0p}), (\ref{Gamma1p}).) 
 
The Faddeev equation satisfied by $\Psi_3$ yields a set of coupled equations 
for the matrix valued functions ${\cal S}$, ${\cal A}_\nu^i$: 
\begin{eqnarray} 
\nonumber\lefteqn{ \left[ \begin{array}{r} 
{\cal S}(k;P)\, u(P)\\ 
{\cal A}^i_\mu(k;P)\, u(P) 
\end{array}\right]}\\ 
&&  = -4\,\int\frac{d^4\ell}{(2\pi)^4}\,{\cal M}(k,\ell;P) 
\left[ 
\begin{array}{r} 
{\cal S}(\ell;P)\, u(P)\\ 
{\cal A}^j_\nu(\ell;P)\, u(P) 
\end{array}\right], 
\label{FEone} 
\end{eqnarray} 
where one factor of ``2'' appears because $\Psi_3$ is coupled symmetrically to 
$\Psi_1$ and $\Psi_2$, and we have evaluated the necessary colour contraction: 
$(H^a)_{bc} (H^a)_{cb^\prime}=-2 \,\delta_{bb^\prime}$. 
 
The kernel in Eq.~(\ref{FEone}) is 
\begin{equation} 
\label{calM} {\cal M}(k,\ell;P) = \left[\begin{array}{cc} 
{\cal M}_{00} & ({\cal M}_{01})^j_\nu \\ 
({\cal M}_{10})^i_\mu & ({\cal M}_{11})^{ij}_{\mu\nu}\rule{0mm}{3ex} 
\end{array} 
\right] 
\end{equation} 
with 
\begin{eqnarray} 
\nonumber \lefteqn{ {\cal M}_{00} = \Gamma^{0^+}(k_q-\ell_{qq}/2;\ell_{qq})\, 
S^{\rm T}(\ell_{qq}-k_q)\,}\\ 
&& \times \,\bar\Gamma^{0^+}(\ell_q-k_{qq}/2;-k_{qq})\, 
S(\ell_q)\,\Delta^{0^+}(\ell_{qq}) \,, 
\end{eqnarray} 
where: $\ell_q=\ell+P/3$, $k_q=k+P/3$, $\ell_{qq}=-\ell+ 2P/3$, 
$k_{qq}=-k+2P/3$; $S$ is the propagator of the dormant dressed-quark 
constituent of the nucleon (Sec.~\ref{subsubsec:S}); and 
\begin{eqnarray} 
\nonumber \lefteqn{({\cal M}_{01})^j_\nu = {\tt t}^j \,
\Gamma_\mu^{1^+}(k_q-\ell_{qq}/2;\ell_{qq})\, 
S^{\rm T}(\ell_{qq}-k_q)\,}\\ 
&&\times \,\bar\Gamma^{0^+}(\ell_q-k_{qq}/2;-k_{qq})\, 
S(\ell_q)\,\Delta^{1^+}_{\mu\nu}(\ell_{qq}) \,,\\ 
\label{calM01} 
\nonumber \lefteqn{({\cal M}_{10})^i_\mu = 
\Gamma^{0^+}(k_q-\ell_{qq}/2;\ell_{qq})\, 
S^{\rm T}(\ell_{qq}-k_q)\,}\\ 
&&\times \,{\tt t}^i\, \bar\Gamma_\mu^{1^+}(\ell_q-k_{qq}/2;-k_{qq})\, 
S(\ell_q)\,\Delta^{0^+}(\ell_{qq}) \,,\\ 
\nonumber \lefteqn{({\cal M}_{11})^{ij}_{\mu\nu} = {\tt t}^j\, 
\Gamma_\rho^{1^+}(k_q-\ell_{qq}/2;\ell_{qq})\, S^{\rm T}(\ell_{qq}-k_q)\,}\\ 
&&\times \,{\tt t}^i\, \bar\Gamma^{1^+}_\mu(\ell_q-k_{qq}/2;-k_{qq})\, 
S(\ell_q)\,\Delta^{1^+}_{\rho\nu}(\ell_{qq}) \,. \label{calM11} 
\end{eqnarray} 
 
In Eqs.~(\ref{FEone})-(\ref{calM11}) it is implicit that $u(P)$ is a normalised 
average of $\varphi_\pm$ so that, e.g., the equation for the proton is obtained 
by projection on the left with $\varphi^\dagger_+$.  To clarify this, by 
illustration, we note that Eq.~(\ref{calM01}) generates an isospin coupling 
between $u(P)_{\varphi_+}$ on the l.h.s.\ of Eq.~(\ref{FEone}) and, on the 
r.h.s., 
\begin{equation} 
\surd 2\,{\cal A}^+_\nu\, u(P)_{\varphi^-} - {\cal A}^0_\nu 
\,u(P)_{\varphi_+}\,. 
\end{equation} 
This is merely the Clebsch-Gordon coupling of 
isospin-$1\oplus\,$isospin-$\frac{1}{2}$ to total isospin-$\frac{1}{2}$ and 
means that the scalar diquark amplitude in the proton, $(ud)_{0^+}\,u$, is 
coupled to itself {\it and} the linear combination: 
\begin{equation} 
\surd 2\, (uu)_{1^+}\, d - (ud)_{1^+} \, u\,. 
\end{equation} 
 
The general forms of ${\cal S}$ and ${\cal A}_\mu^i$, the Bethe-Salpeter-like 
amplitudes that describe the momentum-space correlation between the quark and 
diquark in the nucleon, are discussed at length in Ref.~\cite{oettel}, wherein 
a detailed analysis of the Faddeev equation's solution is presented. Requiring 
that ${\cal S}$ be an eigenfunction of $\Lambda_+(P)$, Eq.~(\ref{Lplus}), 
entails 
\begin{equation} 
{\cal S}(\ell;P) = f_1(\ell;P)\,I_{\rm D} + \frac{1}{M}\left(i\gamma\cdot \ell 
- \ell \cdot \hat P\, I_{\rm D}\right)\,f_2(\ell;P)\,, 
\end{equation} 
where $(I_{\rm D})_{rs}= \delta_{rs}$, $\hat P^2= - 1$, and, in the nucleon 
rest frame, $f_{1,2}$ describe, respectively the upper, lower component of 
the bound-state nucleon's spinor.  Requiring the same of ${\cal A}^i_\mu$ 
reduces to only six (from an original twelve) the number of independent Dirac 
amplitudes required to specify it completely.  However, we simplify this by 
retaining only those two amplitudes that survive in the non-relativistic limit: 
\begin{equation} 
{\cal A}_\mu^i(\ell;P) = a_1^i(\ell;P) \, \gamma_5\gamma_\mu + 
a_2^i(\ell;P)\,\gamma_5 \gamma\cdot\hat\ell \,\hat\ell_\mu\,, \;\hat\ell^2=1. 
\end{equation} 
Assuming isospin symmetry then: $a_j^1=a_j^2=a_j^3$, $j=1,2$. 
 
The Faddeev equation for the nucleon is Eq.~(\ref{FEone}) with the kernel, 
${\cal M}$, given by Eqs.~(\ref{calM})-(\ref{calM11}): to complete its definition 
we must specify the dressed-quark propagator, the diquark Bethe-Salpeter 
amplitudes and the diquark propagators. 
 
\subsubsection{Dressed-quarks} 
\label{subsubsec:S} 
The general form of the dressed-quark propagator is 
\begin{eqnarray} 
S(p) & = & -i \gamma\cdot p\, \sigma_V(p^2) + \sigma_S(p^2)\,, \\ 
     & = & [i\gamma\cdot p\, A(p^2) + B(p^2)]^{-1}\,.\label{SpAB} 
\end{eqnarray} 
It can be obtained by solving the QCD gap equation; i.e., the DSE for the 
dressed-quark self energy, and the many such 
studies~\cite{cdragw,revbasti,revreinhard} yield the model-independent result 
that the wave function renormalisation and dressed-quark mass: 
\begin{equation} 
Z(p^2)=1/A(p^2)\,,\;M(p^2)=B(p^2)/A(p^2)\,, 
\end{equation} 
respectively, exhibit significant momentum dependence for $p^2\lesssim 
1\,$GeV$^2$, which is nonperturbative in origin.  This behaviour was recently 
observed in lattice-QCD simulations~\cite{latticequark}, and 
Refs.~\cite{pmqciv,tandyadelaide} provide quantitative comparisons between 
those results and a modern DSE model.  The infrared enhancement of $M(p^2)$ is 
an essential consequence of dynamical chiral symmetry breaking (DCSB) and is 
the origin of the constituent-quark mass.  With increasing $p^2$ the mass 
function evolves to reproduce the asymptotic behaviour familiar from 
perturbative analyses, and that behaviour is unambiguously evident for $p^2 
\gtrsim 10\,$GeV$^2$~\cite{mr97}. 
 
While numerical solutions of the quark DSE are readily obtained, the utility of 
an algebraic form for $S(p)$ is self-evident.  An efficacious parametrisation 
of $S(p)$, which exhibits the features described above, has been used 
extensively in studies of meson properties~\cite{revbasti,revreinhard} and we 
use it herein.  It is expressed via 
\begin{eqnarray} 
\nonumber \bar\sigma_S(x) & =&  2\,\bar m \,{\cal F}(2 (x+\bar m^2))\\ 
&& + {\cal F}(b_1 x) \,{\cal F}(b_3 x) \, 
\left[b_0 + b_2 {\cal F}(\epsilon x)\right]\,,\label{ssm} \\ 
\label{svm} \bar\sigma_V(x) & = & \frac{1}{x+\bar m^2}\, \left[ 1 - {\cal F}(2 
(x+\bar m^2))\right]\,, 
\end{eqnarray} 
with $x=p^2/\lambda^2$, $\bar m$ = $m/\lambda$, ${\cal F}(x)= [1-\exp(-x)]/x$, 
$\bar\sigma_S(x) = \lambda\,\sigma_S(p^2)$ and $\bar\sigma_V(x) = 
\lambda^2\,\sigma_V(p^2)$.  The mass-scale, $\lambda=0.566\,$GeV, and parameter 
values 
\begin{equation} 
\label{tableA} 
\begin{array}{ccccc} 
   \bar m& b_0 & b_1 & b_2 & b_3 \\\hline 
   0.00897 & 0.131 & 2.90 & 0.603 & 0.185 
\end{array}\;, 
\end{equation} 
were fixed in a least-squares fit to light-meson observables~\cite{mark}. The 
dimensionless $u=d$ current-quark mass in Eq.~(\ref{tableA}) corresponds to 
\begin{equation} 
m=5.1\,{\rm MeV}\,. 
\end{equation} 
($\epsilon=10^{-4}$ in Eq.~(\ref{ssm}) acts only to decouple the large- and 
intermediate-$p^2$ domains.) 
 
The parametrisation expresses DCSB, giving a Euclidean constituent-quark mass 
\begin{equation} 
\label{MEq} M_{u,d}^E = 0.33\,{\rm GeV}, 
\end{equation} 
defined~\cite{mr97} as the solution of $p^2=M^2(p^2)$, whose magnitude is 
typical of that employed in constituent-quark models~\cite{tonyANU,simon} and 
for which the value of the ratio: $M_{u,d}^E/m = 65$, is definitive of 
light-quarks~\cite{mishaSVY}.  In addition, DCSB is also manifest in the vacuum 
quark condensate 
\begin{equation} 
-\langle \bar qq \rangle_0^{1\,{\rm GeV}^2} = \lambda^3\,\frac{3}{4\pi^2}\, 
\frac{b_0}{b_1\,b_3}\,\ln\frac{1\,{\rm GeV}^2}{\Lambda_{\rm QCD}^2} = 
(0.221\,{\rm GeV})^3\,, 
\end{equation} 
where we have used $\Lambda_{\rm QCD}=0.2\,$GeV.  The condensate is 
calculated directly from its gauge invariant definition~\cite{mrt97} after 
making allowance for the fact that Eqs.~(\ref{ssm}), (\ref{svm}) yield a 
chiral-limit quark mass function with anomalous dimension $\gamma_m = 
1$. This omission of the additional $\ln( p^2/\Lambda_{\rm 
QCD}^2)$-suppression that is characteristic of QCD is a practical but not 
necessary simplification. 
 
Motivated by model DSE studies~\cite{entire}, Eqs.~(\ref{ssm}), (\ref{svm}) 
express the dressed-quark propagator as an entire function.  Hence $S(p)$ does 
not have a Lehmann representation, which is a sufficient condition for 
confinement~\cite{fn:confinement}.  Employing an entire function for $S(p)$, 
whose form is only constrained via the calculation of spacelike observables, 
can lead to model artefacts when it is employed directly to calculate 
observables involving large timelike momenta~\cite{ahlig}.  An improved 
parametrisation is therefore being sought.  Nevertheless, no problems are 
encountered for moderate timelike momenta (see, e.g., 
Ref.~\cite{hechtadelaide}) and on the subdomain of the complex plane explored 
in the present calculation the integral support provided by an equally 
efficacious alternative cannot differ significantly from that of our 
parametrisation. 
 
\subsubsection{Diquark Bethe-Salpeter amplitudes} 
\label{subsubsec:Gamma} 
The renormalisation-group-improved rainbow-ladder DSE truncation, employed in 
Refs.~\cite{pieterrho,pieterGamma,mr97}, will yield asymptotic diquark states 
in the strong interaction spectrum.  Such states are not observed and their 
appearance is an artefact of the truncation.  Higher order terms in the 
quark-quark scattering kernel (crossed-box and vertex corrections), whose 
analogue in the quark-antiquark channel do not much affect the properties of 
most of the colour-singlet mesons, act to ensure that QCD's quark-quark 
scattering matrix does not exhibit singularities that correspond to asymptotic 
(unconfined) diquark bound states~\cite{truncscheme}.  Nevertheless, studies 
with kernels that do not produce diquark bound states, do support a physical 
interpretation of the masses obtained using the rainbow-ladder truncation, 
Eq.~(\ref{dqmass}): $m_{qq}$ plays the role of a confined-quasiparticle mass in 
the sense that $l_{qq}=1/m_{qq}$ may be interpreted as a range over which the 
diquark correlation can propagate inside a baryon. These observations motivate 
the {\it Ansatz} for the quark-quark scattering matrix that is employed in 
deriving the Faddeev equation: 
\begin{eqnarray} 
\nonumber 
\lefteqn{ [M_{qq}(k,q;K)]_{rs}^{tu} = }\\ 
&& \sum_{J^P=0^+,1^+,\ldots} \bar\Gamma^{J^P}(k;-K)\, \Delta^{J^P}(K) \, 
\Gamma^{J^P}(q;K)\,. \label{AnsatzMqq} 
\end{eqnarray} 
 
While it is not necessary, one practical means of specifying the $\Gamma^{J^P}$ 
in this equation, which is consistent with the above discussion, is to employ 
the solutions of the ladder-like quark-quark Bethe-Salpeter equation (BSE): 
\begin{eqnarray} 
\nonumber\lefteqn{ \Gamma^{J^P}(k;K) = \int\frac{d^4 q}{(2\pi)^4} \, {\cal 
G}(k-q) \, D_{\mu\nu}^{\rm free}(k-q) }\\ 
&& \nonumber \times \frac{\lambda^a}{2} \gamma_\mu\,S(q+K/2) \, 
\Gamma^{J^P}(q;K) 
\,\left[ \frac{\lambda^a}{2}\gamma_\nu\, S(-q+P/2)\right]^{\rm T}\!, \\ 
&& \label{qqBSE} 
\end{eqnarray} 
where the effective coupling, ${\cal G}(k)$, is calculable using perturbation 
theory for $k^2\gtrsim 1\,$GeV$^2$ and is modelled in the infrared (see, 
e.g., Refs.~\cite{pieterrho,pieterGamma,mr97}), and $D_{\mu\nu}^{\rm 
free}(k)$ is the free gluon propagator.  The amplitude is canonically 
normalised: 
\begin{eqnarray} 
\nonumber 2 \,K_\mu & = & \left[ \frac{\partial}{\partial Q_\mu}\, tr \int 
\frac{d^4 q}{(2\pi)^4}\, \bar\Gamma(q;-K) \, S(q+Q/2) \right. \\ 
&& \times \left.\left.  \rule{0ex}{3ex} \, \Gamma(q;K) \, S^{\rm T}(-q+Q/2) 
\right] \right|_{Q=K}^{K^2=-m_{J^P}^2} \!\!\!\!\!\!\!\!\!\!\!\!\!\!\!\! . 
\label{BSEnorm} 
\end{eqnarray} 
 
Using the properties of the Gell-Mann matrices one finds easily from 
Eq.~(\ref{qqBSE}) that $\Gamma^{J^P}_C:= \Gamma^{J^P}C^\dagger$ satisfies 
exactly the same equation as the $J^P$ colour-singlet meson {\it but} for a 
halving of the coupling~\cite{regdqmass}.  This makes clear that the 
interaction in the $(qq)_{\bar 3_c}$ channel is strong and attractive.  The 
same analysis shows the interaction to be strong and repulsive in the 
$(qq)_{6_c}$ channel. 
 
A complete, consistent solution of Eq.~(\ref{qqBSE}) requires a simultaneous 
solution of the quark-DSE, and while this combined procedure is not 
unmanageable it is a computational 
challenge~\cite{pieterrho,pieterGamma,mr97}.  In addition, we have already 
chosen to simplify our calculations by parametrising $S(p)$, and hence we 
follow Refs.~\cite{jacquesA,jacquesmyriad,cdrqciv,nedm,hechtadelaide} and 
also employ that expedient with $\Gamma^{J^P}\!$, using the following 
one-parameter forms: 
\begin{eqnarray} 
\label{Gamma0p} \Gamma^{0^+}(k;K) &=& \frac{1}{{\cal N}^{0^+}} \, 
H^a\,C i\gamma_5\, i\tau_2\, {\cal F}(k^2/\omega_{0^+}^2) \,, \\ 
\label{Gamma1p} {\tt t}^i \Gamma^{1^+}_\mu (k;K) &=& \frac{1}{{\cal N}^{1^+}}\, 
H^a\,i\gamma_\mu C\,{\tt t}^i\, {\cal F}(k^2/\omega_{1^+}^2)\,, 
\end{eqnarray} 
with the normalisation, ${\cal N}^{J^P}\!$, fixed by Eq.~(\ref{BSEnorm}). Our 
{\it Ans\"atze} retain only that single Dirac-amplitude which would represent a 
point particle with the given quantum numbers in a local Lagrangian density: 
these amplitudes are usually dominant in a BSE 
solution~\cite{conradsep,pieterrho,pieterGamma,mr97,a1b1}. 
 
\subsubsection{Diquark propagators} 
\label{subsub:dqprop} 
Solving for the quark-quark scattering matrix using the ladder-like kernel in 
Eq.~(\ref{qqBSE}) yields free particle propagators for $\Delta^{J^P}$ in 
Eq.~(\ref{AnsatzMqq}).  However, as already noted, higher order contributions 
remedy that defect, eliminating asymptotic diquark states from the spectrum. It 
is apparent in Ref.~\cite{truncscheme} that the attendant modification of 
$\Delta^{J^P}$ can be modelled efficaciously by simple functions that are 
free-particle-like at spacelike momenta but pole-free on the timelike axis. 
Hence we employ~\cite{conradsep} 
\begin{eqnarray} 
\Delta^{0^+}(K) & = & \frac{1}{m_{0^+}^2}\,{\cal F}(K^2/\omega_{0^+}^2)\,,\\ 
\Delta^{1^+}_{\mu\nu}(K) & = & \left(\delta_{\mu\nu} + \frac{K_\mu 
K_\nu}{m_{1^+}^2}\right) \, \frac{1}{m_{1^+}^2}\, {\cal F}(K^2/\omega_{1^+}^2) 
, 
\end{eqnarray} 
where the two parameters $m_{J^P}$ are diquark pseudoparticle masses and 
$\omega_{J^P}$ are the widths characterising $\Gamma^{J^P}$.  It is plain upon 
inspection that these {\it Ans\"atze} satisfy the constraints we have 
elucidated. 
 
\subsection{Model for $\Delta$} \label{subsec:FEDelta} 
The $\Delta$ is a spin-$3/2$, isospin-$3/2$ decuplet baryon and the general 
form of the Faddeev amplitude for such a system is complicated.  However, as 
we assume isospin symmetry, we can focus on the $\Delta^{++}$, with it's simple 
flavour structure, because all the charge states are degenerate.  The Dirac 
structure, though, remains complex and its general form is discussed in 
Ref.~\cite{oettel}.  Herein, as we have for the nucleon, we use that study as 
the guide to a minimal model: 
\begin{equation} 
 \Psi^\Delta_3 = {\tt t}^+ \Gamma^{1^+}_\mu(\frac{1}{2}p_{[12]};K) 
\, \Delta_{\mu\nu}^{1^+}(K) \, \Delta_\nu(\ell;P)\,, \label{DeltaAmpA} 
\end{equation} 
with 
\begin{equation} 
\Delta_\nu(\ell;P) =   {\cal S}^\Delta(\ell;P) \, u_\nu(P)\,\varphi_+ + {\cal 
A}_\nu^\Delta(\ell;P)\, \ell^\perp\cdot u(P)\, \varphi_+ \,,  
\label{DeltaAmpB} 
\end{equation} 
where $u_\nu(P)$ is a Rarita-Schwinger spinor (see Appendix~\ref{App:EM}), 
$\ell^\perp= \ell + \hat P\,\ell\cdot\hat P$, and, again focusing on 
eigenfunctions of $\Lambda_+(P)$, 
\begin{eqnarray} 
\nonumber 
{\cal S}^\Delta(\ell;P) & = & f_1^\Delta(\ell;P)\, I_D \\ 
&& + \frac{1}{M}\left(i\gamma\cdot \ell 
- \ell \cdot \hat P\, I_{\rm D}\right)\,f_2(\ell;P)\,,\\ 
\nonumber {\cal A}_\mu^\Delta(\ell;P) & = & \left[ a_1^\Delta(\ell;P) \,I_D + i 
a_2^\Delta(\ell;P) \,\gamma\cdot\ell^\perp \right] \, \hat P_\mu\,. 
\\ 
& & 
\end{eqnarray} 
The Faddeev equation for the $\Delta$ now assumes the form 
\begin{eqnarray} 
\Delta_\mu(k;P) & = &  4\int\frac{d^4\ell}{(2\pi)^4}\,{\cal 
M}^\Delta_{\mu\nu}(k,\ell;P) \,\Delta_\nu(\ell;P) \label{FEDelta} 
\end{eqnarray} 
with 
\begin{eqnarray} 
\nonumber \lefteqn{{\cal M}^\Delta_{\mu\nu} = {\tt t}^+ 
\Gamma_\rho^{1^+}(k_q-\ell_{qq}/2;\ell_{qq})\, S^{\rm T}(\ell_{qq}-k_q)}\\ 
&& \times \, {\tt t}^+\bar\Gamma^{1^+}_\mu(\ell_q-k_{qq}/2;-k_{qq})\, 
S(\ell_q)\,\Delta^{1^+}_{\rho\nu}(\ell_{qq}) \,. 
\end{eqnarray} 
It is straightforward to construct four projection operators that yield the 
coupled equations for $f^\Delta_{1,2}$, $a^\Delta_{1,2}$.\ 
 
We employ one more expedient to simplify our calculations: we retain only the 
zeroth Chebyshev moments of $f_{1,2}$, $a_{1,2}^i$, $f_{1,2}^\Delta$, 
$a_{1,2}^\Delta$; i.e., we assume $f_1(\ell;P)= f_1(\ell^2;P^2)$, etc.  We 
note that solving integral equations using a Chebyshev decomposition of the 
solution functions is a rapidly convergent scheme for isospin symmetric 
systems~\cite{oettel,mr97,pieterrho,pieterGamma} and neglecting the other 
moments in this calculation will only have a small quantitative effect. 
 
\subsection{Faddeev Equation Masses} 
\label{subsec:FEmass} 
The nucleon and $\Delta$ masses can now be obtained by solving 
Eqs.~(\ref{FEone}), (\ref{FEDelta}), and that also yields the bound-state 
amplitudes necessary for the calculation of the impulse approximation to $N$ 
and $\Delta$ form factors.  The kernels of the equations are constructed from 
the dressed-quark propagator, and the diquark Bethe-Salpeter amplitudes and 
propagators, which are specified in 
Secs.~\ref{subsubsec:S}-\ref{subsub:dqprop}.  These kernels involve four 
parameters.  We fix 
\begin{equation} 
m_{0^+} = 0.74\,{\rm GeV}\,; 
\end{equation} 
i.e., we use the calculated scalar diquark mass in Eq.~(\ref{dqmass}), which is 
consistent with that obtained in recent, more sophisticated BSE 
studies~\cite{pieterdq}. (NB.\ $m_{0^+} \sim 2 M^E$, Eq.~(\ref{MEq}), and hence 
it sets a good scale for nucleon observables.) This leaves $m_{1^+}$ and the 
diquark width parameters, $\omega_{J^P}$.  The immediate goal is to determine 
whether there are intuitively reasonable values of these parameters for which 
one obtains the nucleon and $\Delta$ masses: $M_N = 0.94\,{\rm GeV}$, $M_\Delta 
= 1.23\,{\rm GeV}$, subject to the constraint $m_{1^+}/m_{0^+} \approx 1.3$, as 
in Eq.~(\ref{dqmass}). 
 
\begin{table}[t] 
\caption{\label{tableMass} Calculated nucleon and $\Delta$ masses. The results 
in the first and third rows were obtained using scalar and pseudovector diquark 
correlations: $m_{1^+}=0.90\,$GeV in row 1, $m_{1^+}=0.94\,$GeV in row 3. 
($m_{0^+}=0.74\,$GeV, always.) Pseudovector diquarks were omitted in the second 
and fourth rows.  $\omega_{f_{1,2}}$ are discussed after 
Eq.~(\protect\ref{lqq}), and \mbox{\sc r} in and after Eq.~(\protect\ref{scR}). 
All dimensioned quantities are in GeV.} 
\vspace*{1ex} 
\begin{ruledtabular} 
\begin{tabular*} 
{\hsize} {l@{\extracolsep{0ptplus1fil}} 
|c@{\extracolsep{0ptplus1fil}}c@{\extracolsep{0ptplus1fil}} 
|c@{\extracolsep{0ptplus1fil}}c@{\extracolsep{0ptplus1fil}} 
|c@{\extracolsep{0ptplus1fil}}c@{\extracolsep{0ptplus1fil}}c} 
%
     & $\omega_{0^+}$ & $\omega_{1^+}$ & $M_N$ & $M_\Delta$  & 
     $\omega_{f_1}$ & $\omega_{f_2}$ & {\sc r}\\\hline 
$0^+$ \& $1^+$ & 0.64 & 1.19~ & 0.94 & 1.23~  &  0.49 & 0.44 & 0.25\\ 
$0^+$ & 0.64 & - & 1.59 & -  & 0.39 & 0.41 & 1.28\\\hline 
$0^+$ \& $1^+$ & 0.45 & 1.36~ & 1.14 & 1.33~ & 0.44 & 0.36& 0.54\\ 
$0^+$ & 0.45 & - & 1.44 & -  & 0.36 & 0.35 & 2.32 
\end{tabular*} 
\end{ruledtabular} 
\end{table} 
 
The calculated masses are presented in Table~\ref{tableMass}, from which it
is apparent that the observed masses are easily obtained using solely the
dressed-quark and diquark degrees of freedom we have described above.  The
first two lines of the table also make plain that the additional quark
exchange associated with the introduction of pseudovector correlations
provides considerable attraction.  In this case it reduces the nucleon's mass
by 41\%, in agreement with Ref.~\cite{oettel}, and, of course, without the
$1^+$ correlation the $\Delta$ would not be bound in this approach.
Furthermore, in agreement with intuition, the nucleon and $\Delta$ masses
increase with increasing $m_{J^P}$.
 
The values of the diquark width parameters are reasonable.  For example, with 
\begin{equation} 
\label{lqq} r_p > l_{0^+} := 1/\omega_{0^+} = 0.31\,{\rm fm} > l_{1^+} := 
1/\omega_{1^+} = 0.17\,{\rm fm}\,, 
\end{equation} 
$r_p$ is the proton's charge radius (experimentally, $0.87\,$fm), these 
correlations lie within the nucleon, a point also emphasised by the scalar 
diquark's charge radius, calculated as described in Ref.~\cite{jacquesA}: 
\begin{equation} 
r_{0^+}^2 = (0.55\,{\rm fm})^2 = (0.98\,r_\pi)^2 
\end{equation} 
with $r_\pi$ calculated in the same model~\cite{mark}.  Furthermore, defining
$\omega_{f_{1,2}}$ by requiring a least-squares fit of ${\cal
F}(\ell^2/\omega_{f_{1,2}})$ to $f_{1,2}(\ell^2)$, magnitude-matched at
$\ell^2\simeq 0$, we obtain a scale characterising the quark-diquark
separation
\begin{equation} 
l_{q(qq)_{f_1}}:=1/\omega_{f_1} = 0.40 \, {\rm fm} > 0.15\,{\rm fm} = 
\frac{1}{2}\,l_{0^+}\,. 
\end{equation} 
For the pseudovector analogue 
\begin{equation} 
l_{q(qq)_{a_1}} = 0.36\,{\rm fm} > \frac{1}{2}\,l_{1^+}\,. 
\end{equation} 
($a_2(\ell^2)$ is small in magnitude, slowly varying and not monotonic.  Hence, 
in this case, the fit is of limited use.  Nevertheless, it's momentum space 
width is roughly four times that of $f_1(\ell^2)$.) 
 
For the $\Delta$, 
\begin{equation} 
l_{q(qq)_{f_2^\Delta}}= 0.35\,{\rm fm} \approx l_{q(qq)_{f_1^\Delta}}= 
0.32\,{\rm fm} > \frac{1}{2}\, l_{1^+}\,; 
\end{equation} 
and $a_1^\Delta$ is important, characterised by a peak value of $\approx 
-0.4\,f_1^\Delta(\ell^2=0)$ and $\omega_{a_1^\Delta} \gtrsim 2 
\,\omega_{f_1^\Delta}$, but $a_2^\Delta$ is not: $a_2^\Delta \simeq 0$. 
 
The ratio 
\begin{equation} 
\label{scR} \mbox{\sc r} = f_2(\ell^2=0;-M_N^2)/f_1(\ell^2=0;-M_N^2) 
\end{equation} 
measures the importance of the lower component of the positive energy nucleon's 
spinor and it is not small, which emphasises the importance of treating these 
systems using a Poincar\'e covariant framework.  For the $\Delta$, $\mbox{\sc 
r}=0.17$. 

\section{Pion-induced Nucleon Self-Energy} 
\label{sec:piloop} 
We have illustrated that an internally consistent and accurate description of 
the nucleon and $\Delta$ masses is easily obtained using a Poincar\'e covariant 
Faddeev equation based on confined diquarks and quarks.  However, since the 
$\pi N N$ and $\pi N \Delta$ couplings are large, it is important to estimate 
the shift in the masses due to $\pi$-dressing.  Herein we focus on the shift in 
the nucleon's mass because it is a much studied example. 
 
\subsection{Model Field Theory: Linear Realisation of Chiral Symmetry} 
\label{subsec:modelfieldtheory} 
We begin by considering a model $\pi$-$N$ field theory described by the local 
Lagrangian density: 
\begin{eqnarray} 
\nonumber {\cal L}(x) & = & \bar N(x)\, [ i \! \not\!\partial - M {\cal V}(x) 
]\, N(x) \\ 
\nonumber && + \,\frac{f_\pi^2}{16}{\rm tr} \left[\,\partial_\mu {\cal 
V}^\dagger(x) \, \partial^\mu {\cal V}(x)\right]\\ 
& & - \,\frac{f_\pi^2 m_\pi^2}{16}\,{\rm tr} \left[ 2 - {\cal V}(x) - 
{\cal V}^\dagger(x)\right]\,, 
\label{LpiNNL} 
\end{eqnarray} 
(in this and Sec.~\ref{subsec:modelfieldtheorynonlinear} we employ a Minkowski 
metric) where ``tr'' is a trace over Dirac and isospin indices, $M$ is the 
nucleon's non-pion-dressed mass and the $\pi$-matrix 
\begin{equation} 
{\cal V}(x) = \exp\left(i \gamma_5 \frac{1}{f_\pi} \vec{\tau}\cdot\vec{\pi}(x) 
\right)\,, 
\end{equation} 
with $f_\pi\approx 92\,$MeV, the pion's weak decay constant.  Neglecting the 
$m_\pi^2$- (pion-mass-)term, this Lagrangian exhibits a linear realisation of 
chiral symmetry: 
\begin{eqnarray} 
N(x) & \to &  N^\prime(x) = V(\varphi) \,N(x)\,,\;\\ 
\bar N(x) & \to & \bar N^\prime(x) = \bar N(x) \,V(\varphi)\,, \\ 
{\cal V}(x) & \to & {\cal V}^\prime(x) = V^\dagger(\varphi) {\cal V}(x)\, 
V^\dagger(\varphi)\,, 
\end{eqnarray} 
where $V(\varphi)= \exp\left( i \gamma_5 \vec{\tau}\cdot \vec{\varphi}/f_\pi 
\right)$, with $\vec{\varphi}$ a spacetime-independent three-vector.  (NB.\ The 
form of Eq.~(\ref{LpiNNL}) can be seen to arise from, and express, DCSB at the 
quark level using, e.g., the Global Colour 
Model~\cite{petergcm,gunnergcm,goldstonemode}.  It also arises using the 
rainbow-ladder truncation of the DSEs.) 
 
Using Eq.~(\ref{LpiNNL}) we explore the effect of $\pi$-dressing on $M$ via the 
DSE for the nucleon self energy; i.e., $\Sigma(P)$ in 
\begin{eqnarray} 
G^{-1}(P) & = & \not\! P - M -\Sigma(P)\,. 
\end{eqnarray} 
In rainbow (Hartree-Fock) truncation that equation is: 
\begin{eqnarray} 
\nonumber 
 \Sigma(P) & = & - 3 i g^2\, \int \frac{d^4 k}{(2\pi)^4} \, 
 \Delta(k^2,m_\pi^2)\,\gamma_5\, G(P-k) \, \gamma_ 5 \\ 
& & - 3 i \,\frac{g^2}{2 M} \int \frac{d^4 k}{(2\pi)^4} \, 
\Delta(k^2\,m_\pi^2)\,, \label{SpiNN} 
\end{eqnarray} 
where, from the Lagrangian, 
\begin{equation} 
\label{gAeqone} g= \frac{M}{f_\pi} 
\end{equation} 
(so that $g_A=1$ at tree level in this model) and 
\begin{eqnarray} 
\lefteqn{ \Delta(k^2,m_\pi^2)  = \frac{1}{k^2 - m_\pi^2 + i \varepsilon}}\\ 
& = & \frac{1}{2\,\omega_\pi(\vec{k})}\left[ 
\frac{1}{k_0 - \omega_\pi(\vec{k}) + i\varepsilon} - 
\frac{1}{k_0 + \omega_\pi(\vec{k}) - i\varepsilon} \right] , \nonumber \\ 
&& 
%
\end{eqnarray} 
with $\omega_\pi^2(\vec{k}) = \vec{k}^2 + m_\pi^2$, is the free-pion 
propagator.  The second contribution on the r.h.s.\ in Eq.~(\ref{SpiNN}) is a 
tadpole (Hartree) term, which vanishes if the model is defined via 
dimensional regularisation.  It is generated by the contact term in 
Eq.~(\ref{LpiNNL}): $g^2/(2 M)\, \bar N \vec{\pi} \cdot \vec{\pi} N $, whose 
presence and strength is dictated by chiral symmetry~\cite{adler,sakurai}. 
 
%
As a first step we evaluate the self energy perturbatively.  To proceed with 
that we define the integrals in Eq.~(\ref{SpiNN}) by implementing a 
translationally invariant Pauli-Villars regularisation; i.e., we modify the 
$\pi$-propagator: 
\begin{equation} 
\label{DpiPV} \Delta(k^2,m_\pi^2) \to \bar\Delta_\pi(k^2) 
= \Delta(k^2,m_\pi^2) + \sum_{i=1,2} c_i\, \Delta(k^2,\lambda_i^2)\,, 
\end{equation} 
and then, with 
\begin{equation} 
c_1= - \,\frac{\lambda_2^2-m_\pi^2}{\lambda^2_2-\lambda^2_1}\,,\; 
c_2= \frac{\lambda_1^2-m_\pi^2}{\lambda^2_2-\lambda^2_1} \,,
\end{equation} 
Eq.~(\ref{DpiPV}) yields 
\begin{eqnarray} 
\nonumber \bar\Delta_\pi(k^2) & = & \Delta(k^2,m_\pi^2)\, 
\prod_{i=1,2} \, (\lambda_i^2-m_\pi^2) \, \Delta(k^2,\lambda_i^2)\,,\\ 
&& 
\end{eqnarray} 
in which case the integrals are convergent for any fixed $\lambda_{1,2}$. 
Furthermore, for $m_\pi \ll \lambda_1\to \lambda_2 = \lambda$ 
\begin{eqnarray} 
\label{PVtovertex} 
\bar\Delta_\pi(k^2)& = & \Delta(k^2,m_\pi^2) \, \Delta^2(k^2/\lambda^2,1) 
\end{eqnarray} 
i.e., our Pauli-Villars regularisation is equivalent to employing a monopole 
form factor at each $\pi N N$ vertex: $g \to g\, \Delta(k^2/\lambda^2,1)$, 
where $k$ is the pion's momentum~\cite{fn:PV}.  Since this procedure modifies 
the pion propagator it may be interpreted as expressing compositeness of the 
pion and regularising its off-shell contribution (a related effect is 
identified in Refs.~\cite{reglaws,rhopipipeter}) but that interpretation is not 
unique. 
 
In order to better understand the structure of the self-energy we decompose the 
bare nucleon propagator into a sum of positive and negative energy components 
\begin{eqnarray} 
 G(P) & = & G^+(P) + G^-(P) \\ 
\nonumber & := & \frac{M}{\omega_N(\vec{P})} \left[ \Lambda_+(\vec{P}) 
\frac{1}{P_0 - \omega_N(\vec{P}) + i \varepsilon} \right.\\ 
& & \left.  + \, \Lambda_-(\vec{P}) \frac{1}{P_0 + \omega_N(\vec{P}) - i 
\varepsilon} \right]\,, 
\end{eqnarray} 
where $\omega_N^2(\vec{P}) = \vec{P}^2 + M^2$, and $\Lambda_\pm(\vec{P}) = 
(\not\!\tilde P \pm M)/(2M)$, $\tilde P=(\omega(\vec{P}),\vec{P})$, are, 
respectively, the Minkowski space positive and negative energy projection 
operators.  Now the shift in the mass of a positive energy nucleon is 
\begin{equation} 
\label{defndeltaMp} 
\delta M_+ = \frac{1}{2}{\rm tr}_{D} \left[ \Lambda_+(\vec{P}=0)\, 
\Sigma(P_0=M,\vec{P}=0)\right]\,. 
\end{equation} 
 
We focus initially on the positive-energy nucleon's contribution to the loop 
integral; i.e., the $\Delta(k) \, G^+(P-k)$ contribution in the first term of 
Eq.~(\ref{SpiNN}), which we denote by $\delta_F M_+^+$.  Evaluating the 
$k_0$-integral by closing the contour in the lower half-plane, thereby 
encircling only the three positive-energy pion-like poles, we obtain 
\begin{eqnarray} 
\nonumber \delta_F M_+^+ &= &-3 g^2 \int \frac{d^3 k}{(2\pi)^3} \, 
\frac{\omega_N(\vec{k}^2) - M}{4\,\omega_N(\vec{k}^2)} \\ 
\nonumber && \times \sum_{i=0,1,2}\frac{c_i}{\omega_{\lambda_i}(\vec{k}^2) \, 
[\omega_{\lambda_i}(\vec{k}^2) + \omega_N(\vec{k}^2) - M]} ,\\ 
&&  \label{deltaMpp} 
\end{eqnarray} 
with $c_0=1$, $\lambda_0=m_\pi$ and $\omega_{\lambda_i}^2= 
\vec{k}^2+\lambda_i^2$.  It is obvious that $\delta_F M_+^+<0$; i.e., the Fock 
self-energy diagram's positive energy nucleon piece reduces the mass of a 
positive energy nucleon. 
 
It is instructive to consider Eq.~(\ref{deltaMpp}) further.  Suppose that $M$ 
is very much greater than the other scales then, on the domain in which the 
integrand has significant support, one has 
\begin{equation} 
\omega_N(\vec{k}^2) - M \approx \frac{\vec{k}^2}{2\, M} 
\end{equation} 
and then 
\begin{eqnarray} 
\nonumber \delta_F M_+^+ & \approx & - 3 g^2 \int \frac{d^3 k}{(2\pi)^3} \, 
\frac{\vec{k}^2}{8\,M^2}  \sum_{i=0,1,2} 
\frac{c_i}{\omega_{\lambda_i}^2\!(\vec{k}^2)} \\ 
&& \label{deltaMppNR} 
\end{eqnarray} 
so that 
\begin{eqnarray} 
\frac{d^2 \,\delta_F M_+^+}{(d m_\pi^2)^2} &\approx &-\,\frac{3 g^2}{4 M^2} \, 
\int \frac{d^3 k}{(2\pi)^3} \frac{\vec{k}^2}{\omega_\pi^6(\vec{k}^2)}\\ 
& = & -\, \frac{9}{128 \pi} \frac{g^2}{M^2} \frac{1}{m_\pi}\,. 
\end{eqnarray} 
Thus, on the domain considered, 
\begin{equation} 
\label{LNA} \delta_F M_+^+ = -\,\frac{3}{32\pi} \frac{g^2}{M^2} m_\pi^3 + 
f^+_{(1)}(\lambda_1,\lambda_2)\,m_\pi^2 + f^+_{(0)}(\lambda_1,\lambda_2)\,, 
\end{equation} 
where, as the derivation makes transparent, $f_{(0,1)}$ are scheme-dependent 
functions of (only) the regularisation parameters but the first term is 
regularisation-scheme-independent.  This first term is nonanalytic in the 
current-quark mass and its coefficient is fixed by chiral symmetry.  (NB.\ If 
$\lambda_{1,2}$ are interpreted as setting a compositeness scale for the $\pi 
N N$ vertex, and assume soft values; 
e.g.~\cite{jacquesmyriad,oettel2,tonysoft}, $\sim 600\,$MeV, then the 
quantitative value of $\delta_F M_+^+$ is completely determined by the 
regularisation-scheme-dependent terms.) 
 
We turn now to the $\Delta(k)\, G^-(P-k)$ contribution in the first term of 
Eq.~(\ref{SpiNN}), which we denote by $\delta_F M_+^-$.  This describes the 
Z-diagram (anti-nucleon) contribution to the nucleon's mass and it is most 
efficient in this case to close the $k_0$-integration contour in the 
upper-half plane, thereby encircling only the three negative-energy pion-like 
poles: 
\begin{eqnarray} 
\nonumber \delta_F M_+^- & = & 3 g^2 \int \frac{d^3 k}{(2\pi)^3} \, 
\frac{\omega_N(\vec{k}^2) + M}{4\,\omega_N(\vec{k}^2)} \\ 
\nonumber &&\times \sum_{i=0,1,2}\frac{c_i}{\omega_{\lambda_i}(\vec{k}^2) \, 
[\omega_{\lambda_i}(\vec{k}^2) + \omega_N(\vec{k}^2) + M]} . \\ 
&& \label{deltaMpm} 
\end{eqnarray} 
It is obvious that $\delta_F M_+^- > 0$; i.e., the Fock diagram's anti-nucleon 
contribution to the positive-energy nucleon's mass is positive, and it is 
equally clear that, as evaluated with a pseudoscalar coupling~\cite{cdrqciv}, 
\begin{eqnarray} 
\nonumber  \delta_F M_+^+ +  \delta_F M_+^- & = & 3 g^2 \int \frac{d^3 
k}{(2\pi)^3} \, 
\frac{M}{2\,\omega_N(\vec{k}^2)} \\ 
\nonumber &&\times \sum_{i=0,1,2} 
\frac{c_i}{[\omega_{\lambda_i}(\vec{k}^2) + 
\omega_N(\vec{k}^2)]^2 - M^2} ;\\ 
& & \label{deltaMp} 
\end{eqnarray} 
i.e., $\delta_F M_+^+ + \delta_F M_+^- > 0$, and hence that the negative-energy 
nucleon contribution overwhelms that of the positive-energy nucleon.  If 
$\delta_F M_+ >0$ were the final word on the mass shift it would contradict all 
previous results for the effect of pion loops on the nucleon's 
mass~\cite{regwrong}. 
 
Before addressing this issue we note that for $M$ very much greater than the 
other scales then, reapplying the analysis that led to Eq.~(\ref{LNA}), 
\begin{eqnarray} 
\nonumber  \delta_F M_+^- & = & f^-_{(0)}(\lambda_1,\lambda_2) + m_\pi^2\, 
f^-_{(1)}(\lambda_1,\lambda_2) \\ 
&& + \,\frac{3}{32\pi^2} \frac{g^2}{M}\, m_\pi^2\, (\ln m_\pi^2 - 1)\,. 
\end{eqnarray} 
The last term on the r.h.s.\ of this equation is an additional nonanalytic 
contribution to the nucleon's mass, and it is of lower order in $1/M$ than the 
nonanalytic term in Eq.~(\ref{LNA}).  This result, if it were to remain 
unameliorated, would also be in conflict with modern theory. 
 
\begin{figure}[t] 
\centerline{\includegraphics[height=6.5cm]{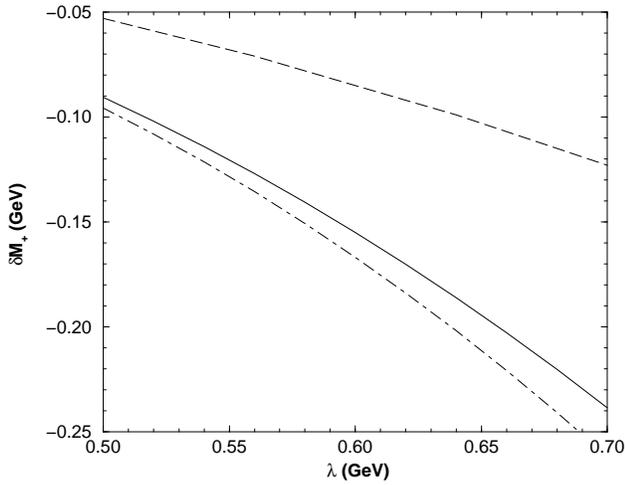}} 
\caption{\label{shiftmassg2} Solid line: Shift in a positive-energy nucleon's 
mass due to the O$(g^2)$ $\pi$-contribution to the self energy, 
Eq.~(\protect\ref{deltaMg2}), obtained using a soft monopole pion-nucleon form 
factor to regularise the pion's off-shell behaviour. ($M=0.94\,$GeV, 
$m_\pi=0.14\,$GeV and $g_A=1$.)  $\delta M_+(\lambda= 0.6\,{\rm GeV})= 
-0.15\,$GeV. Dashed line: $\delta_F M_+^+$; dot-dashed line: $\delta^A M_+^+$.} 
\end{figure}

Hitherto we have neglected the last term in Eq.~(\ref{SpiNN}), which 
describes the tadpole diagram's contribution to the positive-energy nucleon's 
mass shift, and the resolution of these apparent conflicts lies here.  It is 
easy to evaluate and for $\lambda_1 \to \lambda_2 = \lambda$ 
\begin{equation} 
\delta_H M_+ = -\, \frac{3}{32 \pi^2} \frac{g^2}{M} \, \left[ \lambda^2 + 
m_\pi^2 \left( \ln\left[m_\pi^2/\lambda^2\right] - 1 \right) \right]\,. 
\end{equation} 
The inclusion of $\delta_H M_+$ solves both problems.  It provides for an 
exact, algebraic cancellation of the order-$1/M$ term in $\delta_F M_+^-$ that 
is nonanalytic in the current-quark mass, thereby ensuring that the nonanalytic 
term in Eq.~(\ref{LNA}) provides the leading O$(1/M)$ contribution to the 
nucleon's mass.  The cancellation occurs because the $(1/M)^1$ contribution 
from the Z-diagram has the structure of a tadpole term, for reasons that are 
intuitively obvious given that the $(1/M)$-expansion begins with an infinitely 
heavy nucleon.  Furthermore, it must be exact because using dimensional 
regularisation, for example, all tadpole terms vanish and the leading 
nonanalytic term must be regularisation-scheme-independent.  In addition, for 
all $\lambda>0$, 
\begin{equation} 
\label{deltaMg2} \delta M_+ = \delta_F M_+^+ + \delta_F M_+^- + \delta_H M_+ < 
0\,; 
\end{equation} 
i.e., the pion loop reduces the nucleon's mass.  (NB.\ $\delta M_+ (\lambda)$ 
decreases monotonically from $0$ with increasing $\lambda$, see 
Fig.~\ref{shiftmassg2}.) 
 
We opened by asking for the scale of the mass shift produced by the pion loop. 
If we allow the interpretation of the Pauli-Villars regularisation procedure as 
introducing a monopole form factor at each $\pi NN$ vertex, which modifies the 
pion's off-shell behaviour, then using soft values of the monopole scale: 
$\lambda \sim 0.5$-$0.7\,$GeV, as determined in quark-diquark 
Faddeev-amplitude models of the nucleon~\cite{jacquesmyriad,oettel2} and 
inferred from data~\cite{tonysoft}, the O$(g^2)$ shift is as depicted in 
Fig.~\ref{shiftmassg2}. The magnitude is that of Refs.~\cite{tonyCBM,ishii}. 
However, it is evident, and important to note, that this magnitude is extremely 
sensitive to the monopole's scale: centred on $\lambda=0.6\,$GeV, a 10\% change 
in $\lambda$ produces a 30\% change in $\delta M_+$. 
 
\subsection{Nonlinear Realisation of Chiral Symmetry} 
\label{subsec:modelfieldtheorynonlinear} 
An alternative to Eq.~(\ref{LpiNNL}) is to build a Lagrangian density that
contains only derivatives of the pseudoscalar field and thereby expresses a
nonlinear realisation of chiral symmetry~\cite{weinberg}.  In chiral quark
models such a Lagrangian can be obtained via a unitary transformation of the
fields in Eq.~(\ref{LpiNNL}) to obtain a so-called volume (pseudovector)
coupling~\cite{goldstonemode,Thomas:1981ps}.  The leading term in the
nonlinear chiral Lagrangian can easily be obtained by using the equations of
motion for a free nucleon to re-express Eq.~(\ref{LpiNNL}).  Neglecting that
part of the Lagrangian density which describes the pseudoscalar field alone,
this procedure yields:
\begin{equation} 
\label{LpiNL} \bar N(x)\left[ i \not \!\partial - M  +\frac{g}{2 M} \gamma_5 
\gamma^\mu \,\vec{\tau} \cdot \partial_\mu\vec{\pi}(x) + \ldots \right] N(x)\,, 
\end{equation} 
and the rainbow truncation of this model's DSE is 
\begin{eqnarray} 
\nonumber \Sigma(P) & = & 3i \frac{g^2}{4 M^2} \int \frac{d^4 k}{(2\pi)^4} \, 
 \Delta(k^2,m_\pi^2) \! \not\! k\,\gamma_5\, G(P-k)  \! \not\! k\,\gamma_ 5\,. \\ 
&&\label{AVDSE} 
\end{eqnarray} 
No interaction survives that can generate a tadpole (Hartree) term. 
 
We again evaluate this self energy as a one-loop correction to the 
positive-energy nucleon's mass.  The contribution of the positive-energy 
nucleon is 
\begin{eqnarray} 
\nonumber \lefteqn{\delta^A M_+^+  =  -\,\frac{3 g^2}{16 M^2} \int\frac{d^3 
k}{(2\pi)^3} \frac{1}{\omega_N} } \\ 
&& \times \sum_{i=0,1,2} c_i\, \frac{\lambda_i^2 (\omega_N - M) + 2 \vec{k}^2 
(\omega_{\lambda_i} + \omega_N)} {\omega_{\lambda_i} [ \omega_{\lambda_i} + 
\omega_N - M ]}\, , 
\label{deltaMApp} 
\end{eqnarray} 
with $\omega_N=\omega_N(\vec{k}^2)$, etc.  Now, to make transparent the 
direct connection between our approach and other mass-shift calculations, we 
rewrite Eq.~(\ref{deltaMApp}) in the form 
\begin{eqnarray} 
\nonumber \lefteqn{\delta^A M_+^+ = }\\ 
& & \nonumber -\,6\pi\, \frac{f^2_{NN\pi}}{m_\pi^2} \int\frac{d^3 
k}{(2\pi)^3} \, \frac{\vec{k}^2\,u^2(\vec{k}^2)}{\omega_\pi(\vec{k}^2) [ 
\omega_\pi(\vec{k}^2) + \omega_N(\vec{k}^2) - M]} \,,\\ 
&&  \label{deltaMppCBM} 
\end{eqnarray} 
where, as usual, $f_{NN\pi}^2 = g^2 m_\pi^2/(16 \pi M^2)$ and, obviously, 
\begin{eqnarray} 
\vec{k}^2\,\nonumber u^2(\vec{k}^2) & = &
\frac{\omega_{\lambda_0}}{2\,\omega_N} \,  
[ \omega_{\lambda_0} + \omega_N - M ] \\ 
&& \nonumber  \times 
\sum_{i=0,1,2} c_i\, \frac{\lambda_i^2 (\omega_N - M) + 2 \vec{k}^2 
(\omega_{\lambda_i} + \omega_N)} {\omega_{\lambda_i} [ \omega_{\lambda_i} + 
\omega_N - M ]}\,. \\ 
&& \label{ukdef} 
\end{eqnarray} 
This is useful because, for $m_\pi \ll  \lambda_1 \to \lambda_2= 
\lambda$; i.e., on the domain in which Eq.~(\ref{PVtovertex}) is valid, one 
finds algebraically that 
\begin{equation} 
\label{uklimit} 
u(\vec{k}^2) = 1/(1+\vec{k}^2/\lambda^2)\,, 
\end{equation} 
which firmly establishes the qualitative equivalence between 
Eq.~(\ref{deltaMApp}) and the calculation in 
Refs.~\cite{tonyANU,tonyCBM,harry}. 
 
\begin{figure}[t] 
\centerline{\includegraphics[height=6.5cm]{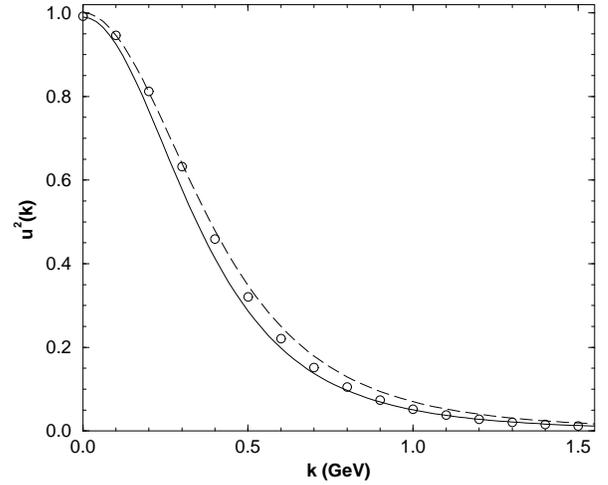}} \caption{\label{plotuk} 
Illustrating that Pauli-Villars regularisation with finite mass-scales is a 
practical tool.  Open circles - $u^2(\vec{k})$ calculated directly from 
Eq.~(\protect\ref{ukdef}) using $m_\pi=0.14\,$GeV, $M=0.94\,$GeV, $g_A=1$ and 
$\lambda_1\to\lambda_2 = \lambda = 0.6\,$GeV; solid line - least-squares fit to 
$\vec{k}^2 u^2(\vec{k}^2)$, which yields: 
$u(\vec{k}^2)=\mbox{$0.99/(1+\vec{k}^2/\bar\lambda^2)^2$}$, $\bar\lambda= 
0.54\,$GeV; dashed line - limiting form from Eq.~(\protect\ref{uklimit}).} 
\end{figure}

In Fig.~\ref{plotuk} we compare the limiting form, Eq.~(\ref{uklimit}), with 
$u(\vec{k})$ calculated from Eq.~(\ref{ukdef}).  This emphasises the practical 
utility of using a Pauli-Villars regularisation to represent a $\pi N N$ vertex 
form factor. 
 
To provide a quantitative connection with other analyses we employ 
\begin{equation} 
\label{CBMuk} u(\vec{k}) = 3 \, \frac{j_1(|\vec{k}| R)}{|\vec{k}| R}\,; 
\end{equation} 
in Eq.~(\ref{deltaMppCBM}); i.e., the CBM form for $u(k)$, where $R$ is the bag 
radius and $j_1(x)$ is a spherical Bessel function.  The results are given in 
Table~\ref{CBMcalc} and may be summarised as (in GeV) 
\begin{equation} 
\label{CBMshift} -\delta^A_{\rm CBM} M_+^+=  (0.065\pm 0.022) \,g_A^2 \,, 
\end{equation} 
where $g_A$ is the nucleon's axial vector coupling constant.  (NB.\ The result 
in Eq.~(\ref{CBMshift}) is also that obtained using a monopole form factor with 
the very soft scale $\lambda_{\rm CBM} = 0.38 \pm 0.04\,$GeV.) We stress that 
in Eqs.~(\ref{LpiNNL}) and (\ref{LpiNL}) we used the coupling $g=M/f_\pi$, 
Eq.~(\ref{gAeqone}), which corresponds to $g_A=1$, whereas using the 
experimental value, $g_A=1.26$, Eq.~(\ref{CBMshift}) gives 
\begin{equation} 
\label{CBMGeVshift} \delta^A_{\rm CBM} M_+^+ = - 0.104 \pm 0.035\,{\rm GeV}\,. 
\end{equation} 
The larger shift described in Ref.~\cite{tonyANU,tonyCBM} is obtained from 
Eq.~(\protect\ref{deltaMppCBM}) by using a smaller bag radius ($\sim 
0.75\,$fm), which is needed to describe $\pi N$ scattering.  The value of $R$ 
employed herein is appropriate to the calculation of nucleon electromagnetic 
form factors~\cite{tonyR}.  \textit{A priori} it is not clear which should be 
used for the calculation of hadron masses but recent lattice 
studies~\cite{Young:2001nc} favour a harder value. 
 
\begin{table}[t] 
\caption{\label{CBMcalc} $\delta^A M_+^+$ calculated using 
Eq.~(\protect\ref{CBMuk}) in Eq.~(\protect\ref{deltaMppCBM}); i.e., a CBM 
estimate. The optimal bag radius for a description of the neutron's electric 
form factor is $R=0.95\,{\rm fm}=1/(0.21\,{\rm GeV})$~\protect\cite{tonyR}.} 
\vspace*{1ex} 
\begin{ruledtabular} 
\begin{tabular*} 
{\hsize} {c@{\extracolsep{0ptplus1fil}} 
|d@{\extracolsep{0ptplus1fil}}d@{\extracolsep{0ptplus1fil}}d} 
%
$R$ (fm) & 0.85 & 0.95 & 1.05 \\ 
$-\delta^A M_+^+$~ (GeV)  & 0.091 & 0.065 & 0.048 
\end{tabular*} 
\end{ruledtabular} 
\end{table} 
 
Returning to Eq.~(\ref{AVDSE}), the mass-shift contribution from the 
negative-energy nucleon; i.e., the Z-diagram, is 
\begin{eqnarray} 
\nonumber \lefteqn{\delta^A M_+^- =  \frac{3 g^2}{16 M^2} \int\frac{d^3 
k}{(2\pi)^3} \frac{1}{\omega_N} } \\ 
&& \times \sum_{i=0,1,2} c_i\, \frac{\lambda_i^2 (\omega_N + M) + 2 \vec{k}^2 
(\omega_{\lambda_i} + \omega_N)} {\omega_{\lambda_i} [ \omega_{\lambda_i} + 
\omega_N + M ]} . \label{dAMpm} 
\end{eqnarray} 
In this case we have $\delta^A M_+^+ < 0$ and $\delta^A M_+^- > 0$ but the sum 
\begin{eqnarray} 
 \delta^A M_+ & = & \delta^A M_+^+ + \delta^A M_+^- \\ 
\nonumber 
 & = & -\,\frac{3 g^2}{8 M^2} 
\int\frac{d^3 k}{(2\pi)^3} \frac{M}{\omega_N}\\ 
&& \times \sum_{i=0,1,2} c_i\, \frac{ 2\vec{k}^2\, 
(\omega_{\lambda_i} + \omega_N) - \lambda_i^2 
\,\omega_{\lambda_i}}{\omega_{\lambda_i}\, (\omega_N + \omega_{\lambda_i})^2 
- M^2} 
\end{eqnarray} 
is self-evidently negative; i.e., with a pseudovector coupling the Z-diagram is 
much suppressed. 
 
Considering the heavy-nucleon limit again one obtains 
\begin{equation} 
\label{ALNA} \delta^A M_+^+  = -\,\frac{3}{32\pi} \frac{g^2}{M^2} m_\pi^3 + 
f^+_{(1^A)}(\lambda_1,\lambda_2)\,m_\pi^2 + f^+_{(0^A)}(\lambda_1,\lambda_2)\,; 
\end{equation} 
i.e., the same contribution, nonanalytic in the current-quark mass, as in 
Eq.~(\ref{LNA}), but with different regularisation-dependent terms.  In this 
case, however, because the Z-diagrams are suppressed by the pseudovector 
coupling, the leading-order contribution to $\delta^A M_+^-$ is O$(1/M)^3$. 
This is clear from Eq.~(\ref{dAMpm}), and makes immediately unambiguous the 
origin and nature of the leading-order nonanalytic contribution to the 
nucleon's mass. 
 
Again interpreting the Pauli-Villars regularisation as introducing a monopole 
form factor at each $\pi N N$ vertex, we can estimate the magnitude of the 
$\pi$-loop's contribution to the nucleon's mass.  Our results are depicted in 
Fig.~\ref{shiftmassg2}.  It is evident that $\delta^A M_+^+ \neq \delta_F 
M_+^+$, which illustrates the difference between the regularisation-dependent 
terms in Eqs.~(\ref{LNA}) and (\ref{ALNA}).  In addition, although it may not 
be immediately obvious, 
\begin{equation} 
\label{ONSHELL} \delta^A M_+ \equiv \delta M_+\,, 
\end{equation} 
which is why there is only one solid curve in the figure.  This result provides 
a quantitative verification of the on-shell equivalence of the pseudoscalar and 
pseudovector interactions, in perturbation theory, as long as the pseudoscalar 
interaction is treated in a manner consistent with chiral 
symmetry~\cite{sakurai}. It also emphasises that, at least for estimating the 
mass shift, it is advantageous to employ the pseudovector interaction.  We 
note, however, that in fully embracing a Lagrangian density that expresses a 
nonlinear realisation of chiral symmetry one loses a direct correspondence with 
extant, ordered truncations of the DSEs, and hence also loses this 
correspondence between the Lagrangian's degrees of freedom and hadrons as 
composites of dressed-quarks. 
 
\subsection{Model DSE} 
\label{subsec:modelDSE} 
We now build on the above analysis and seek a nonperturbative estimate of the
$\pi$-loop's contribution to the nucleon's mass.  Returning to the Euclidean
metric described in Appendix~\ref{App:EM}, which is advantageous for
numerical studies, the DSE for the nucleon's self energy using a
\textit{pseudovector} coupling is
\begin{eqnarray} 
\nonumber \Sigma(P) &= &  3 \int\!\frac{d^4 k}{(2\pi)^4}\, 
g_{PV}^2(P,k) \,\Delta_\pi((P-k)^2)\, \\ 
&& \times \,\gamma\cdot (P-k)\gamma_5\, G(k)\, \gamma\cdot (P-k)\gamma_5\,, 
\label{EDSE} 
\end{eqnarray} 
with the following equivalent representations for the nucleon propagator: 
\begin{eqnarray} 
G(k) & = & 1/[i \gamma\cdot k + M + \Sigma(P)] \,,\\ 
& = & 1/[i \gamma\cdot k \, {\cal A}(k^2) + M + {\cal B}(k^2)]\,, \\ 
& = & -i \gamma\cdot k \,\sigma_{\cal V}(k^2) + \sigma_{\cal S}(k^2)\,, 
\end{eqnarray} 
where $M$ is the nucleon's bare mass, which is obtained, e.g., by solving the
Faddeev equation.  In Eq.~(\ref{EDSE}), $\Delta_\pi(k^2)=1/[k^2+m_\pi^2]$ is
the pion propagator, and $g_{PV}(P,k)$ is a form factor that we will use to
describe the composite nature of {\it both} the pion and the nucleon.  The
self-consistent solution of Eq.~(\ref{EDSE}) yields ${\cal A}(k^2)$ and
${\cal B}(k^2)$, and thereby the nonperturbative mass shift.  (For clarity we
omit a discussion of renormalisation but remark on its effects following
Eq.~(\protect\ref{shiftF}).)
 
We now turn to the model specified by 
\begin{equation} 
\label{gPVexp} g_{PV}(P,k) = \frac{g}{2 M}\,\exp(-(P-k)^2/\Lambda^2)\,. 
\end{equation} 
The exponential form facilitates an algebraic evaluation of many necessary 
integrals and, as has been observed elsewhere~\cite{HLold}, is 
phenomenologically equivalent to a monopole form factor: 
$1/(1+(P-k)^2/\lambda^2)$, if the mass-scales are related via $\Lambda \approx 
\sqrt{2}\,\lambda$.  Thus one can anticipate a quantitative correspondence 
between the $\lambda=0.6\,$GeV monopole results of the previous subsections and 
those obtained in this with $\Lambda \approx 0.9\,$GeV. 
 
\begin{figure}[t] 
\centerline{\includegraphics[height=6.5cm]{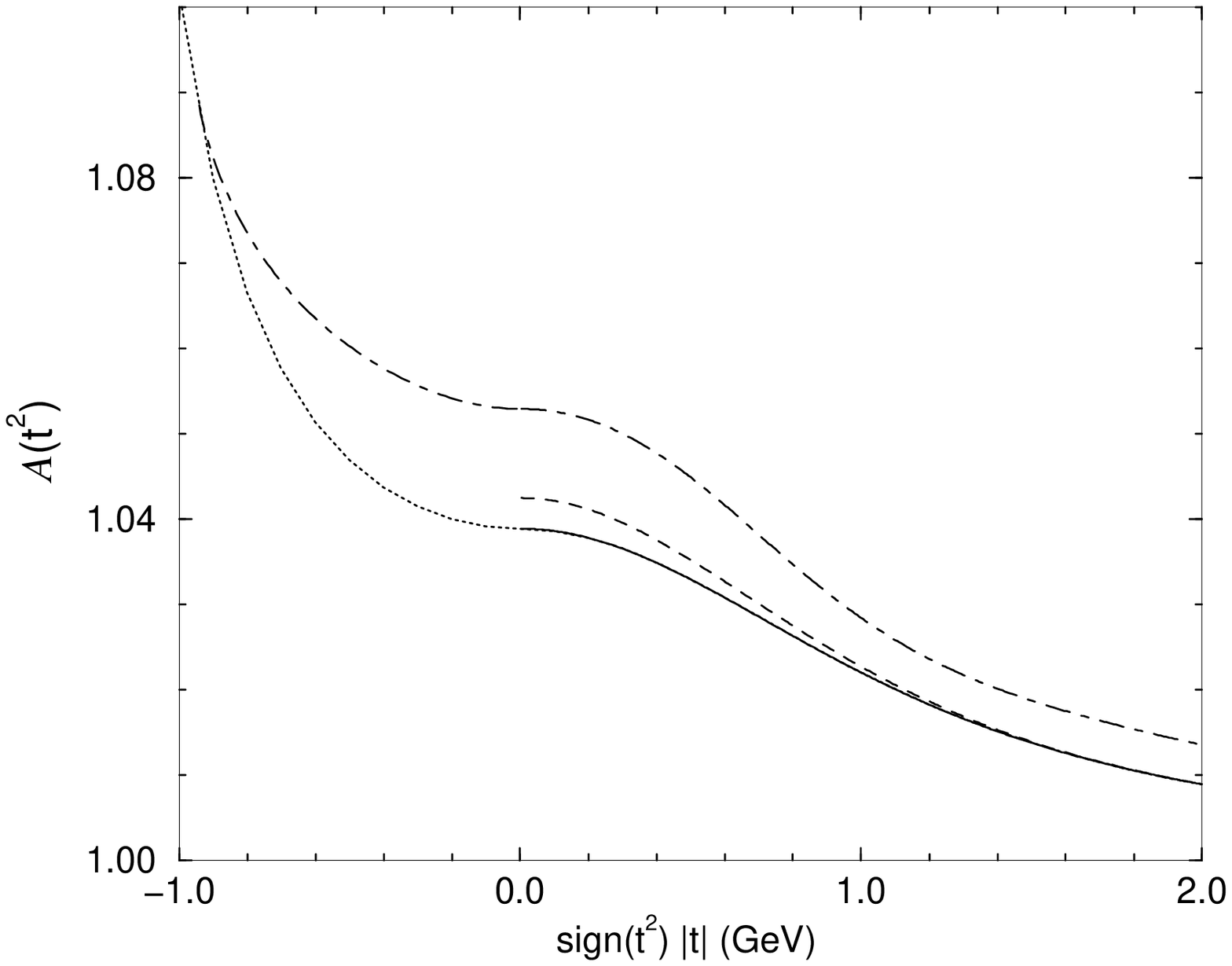}} \caption{\label{cfA} 
Vector piece of the inverse dressed-nucleon propagator.  Dotted line: ${\cal 
A}(t)$ from Eq.~(\protect\ref{AEoneloop}), ${\cal A}(t=4)=1.001$; solid line: 
numerical result for the one-loop-dressed function in the spacelike region, 
obtained from Eqs.~(\protect\ref{KAang}) - (\protect\ref{BEDSE}), which 
overlies the dotted line in this region; dashed line: ${\cal A}(t)$ obtained in 
the self-consistent solution of Eqs.~(\protect\ref{AEDSE}), 
(\protect\ref{BEDSE}); dot-dashed line: ${\cal A}(t)$ (${\cal A}(t=4)=1.002$) 
obtained in the self-consistent solution of Eqs.~(\protect\ref{AEDSEapp}), 
(\protect\ref{BEDSEapp}) with Eqs.~(\protect\ref{nonanA}), 
(\protect\ref{nonanB}) added in the continuation to the timelike region. (All 
curves obtained with $M=0.94\,$GeV, $m_\pi=0.14\,$GeV, $g_A=1$, 
$\Lambda=0.9\,$GeV.)} 
\end{figure}

\begin{figure}[t] 
\centerline{\includegraphics[height=6.5cm]{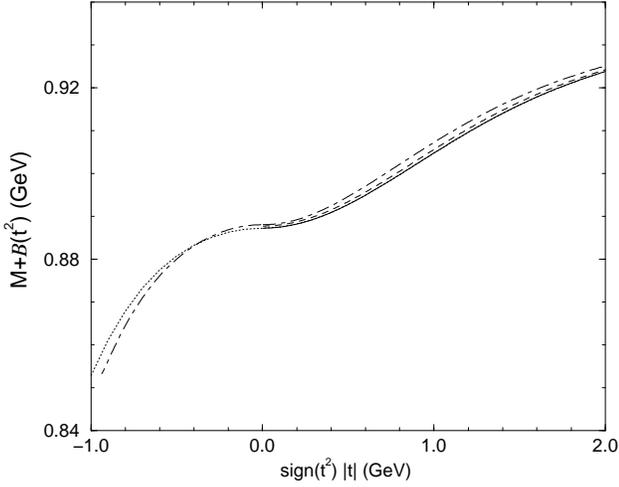}} \caption{\label{cfB}
Scalar piece of the inverse dressed-nucleon propagator.  Dotted line: ${\cal
B}(t)$ from Eq.~(\protect\ref{BEoneloop}), $M+{\cal B}(t=4)=0.937\,$GeV;
solid line: numerical result for the one-loop-dressed function in the
spacelike region, obtained from Eqs.~(\protect\ref{KAang}) -
(\protect\ref{BEDSE}), which overlies the dotted line in this region; dashed
line: ${\cal B}(t)$ obtained in the self-consistent solution of
Eqs.~(\protect\ref{AEDSE}), (\protect\ref{BEDSE}); dot-dashed line: ${\cal
B}(t)$ obtained in the self-consistent solution of
Eqs.~(\protect\ref{AEDSEapp}), (\protect\ref{BEDSEapp}) with
Eqs.~(\protect\ref{nonanA}), (\protect\ref{nonanB}) added in the continuation
to the timelike region. (All curves obtained with $M=0.94\,$GeV,
$m_\pi=0.14\,$GeV, $g_A=1$, $\Lambda=0.9\,$GeV.)}
\end{figure}

Before proceeding with a nonperturbative solution of the nucleon's DSE we 
evaluate the one-loop self-energy so as to provide a direct Euclidean space 
comparison with Secs.~\ref{subsec:modelfieldtheory}, 
\ref{subsec:modelfieldtheorynonlinear}. Using Eq.~(\ref{gPVexp}) we can 
evaluate the $k_4$ integral to obtain 
\begin{eqnarray} 
\nonumber \lefteqn{ {\cal A}(t^2) - 1} \\ 
& = & - \frac{3}{32 \pi^2} \frac{g^2}{M^2} 
\int_0^\infty d\kappa\,\kappa^2\, 
\frac{a(t,\kappa)\,{\rm e}^{-2\kappa^2/\Lambda^2}}{\omega_\pi(\kappa) \, 
\omega_N(\kappa)}, \label{AEoneloop}\\ 
{\cal B}(t^2) & = & 
\label{BEoneloop} - \frac{3}{32 \pi^2} \frac{g^2}{M^2} \int_0^\infty 
d\kappa\,\kappa^2\, \frac{b(t,\kappa)\,{\rm 
e}^{-2\kappa^2/\Lambda^2}}{\omega_\pi(\kappa) \, \omega_N(\kappa)}, 
\end{eqnarray} 
where $a(t,\kappa)$, $b(t,\kappa)$ are given in Eqs.~(\ref{a0tk})-(\ref{a2tk}). 
${\cal A}$ and ${\cal B}$ are plotted in Figs.~\ref{cfA}, \ref{cfB}. 
 
\begin{figure}[t] 
\centerline{\includegraphics[height=6.5cm]{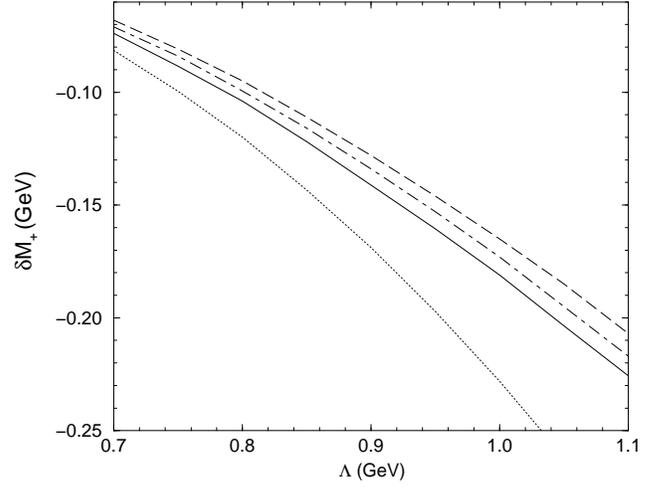}} \caption{\label{Eshift} 
Dashed line: nucleon's one-loop mass-shift, calculated from ${\cal A}$, ${\cal 
B}$ in Eqs.~(\protect\ref{AEoneloop}), (\protect\ref{BEoneloop}) - $\delta 
M_+(\Lambda=0.9\,{\rm GeV}\,\approx \sqrt{2}\lambda)=-0.13\,$GeV; dot-dashed 
line: one-loop mass shift obtained using the approximate kernels in 
Eqs.~(\protect\ref{KAapprox}), (\protect\ref{KBapprox}); solid line: mass shift 
obtained via the self-consistent solution of the nucleon's DSE using these 
approximate kernels - $\delta M_+(\Lambda=0.9\,{\rm GeV})=-0.14\,$GeV.  The 
dotted line is $\delta^A M_+^+$; i.e., Eq.~(\protect\ref{deltaMApp}) calculated 
in our Euclidean model. (All curves obtained with $M=0.94\,$GeV, $m_\pi = 0.14 
\,$GeV, $g_A=1$.)} 
\end{figure}

The one-loop-corrected nucleon mass: $M_{D^1}$, is the solution of 
\begin{equation} 
M_{D^1}^2 {\cal A}^2(-M_{D^1}^2) = [M + {\cal B}(-(M_{D^1}^2)]^2\,, 
\end{equation} 
and it is straightforward to show that $M_{D^1} - M\equiv \delta M_+$, where
$\delta M_+$ is defined in Eq.~(\ref{defndeltaMp}).  The calculated
$\Lambda$-dependence of $\delta M_+$ is depicted in Fig.~\ref{Eshift}, and a
comparison with Fig.~\ref{shiftmassg2} reveals the equivalence between the
Minkowski and Euclidean space formulations.
 
The new feature in a nonperturbative study is that the position of the pole in 
the nucleon's propagator is not known {\it a priori}: locating it is the goal, 
and this precludes an algebraic evaluation of the $k_4$-integral.  The position 
of the pole will depend on the strength of the interaction and the nature of 
the form factor.  In this case one must proceed by first evaluating the angular 
integrals in Eq.~(\ref{EDSE}), which are independent of $G(k)$, noting that for 
a given function of $(P-k)^2$: 
\begin{equation} 
\int d\Omega_k \,f((P-k)^2)= \frac{2}{\pi}\int_{-1}^{1}\! dz\sqrt{1-z^2}\, 
f(P^2+k^2-2 P k z)\,. 
\end{equation} 
This yields the kernels of the coupled, nonlinear integral equations for ${\cal 
A}$, ${\cal B}$: 
\begin{eqnarray} 
\nonumber {\cal K}_{\cal A}(P^2,k^2) &= & \nonumber \frac{1}{2}\int\! 
d\Omega_k\, 
g_{PV}^2((P-k)^2) \bigg[-(P^2 + k^2)  \\ 
& &  \,  + \, 
\frac{(P^2-k^2)^2+m_\pi^2 (P^2+k^2)}{(P-k)^2 + m_\pi^2} \bigg] \,,
\label{KAang}\\  
\nonumber {\cal K}_{\cal B}(P^2,k^2) & = & \int\! d\Omega_k\,
g_{PV}^2((P-k)^2) \\  
& & \times  \bigg[1 - \frac{2 m_\pi^2}{(P-k)^2 + m_\pi^2} \bigg] \, ,
\label{KBang} 
\end{eqnarray} 
so that for spacelike $P^2$ these integral equations can be written ($x=P^2$, 
$y=k^2$) 
\begin{eqnarray} 
\nonumber  x [{\cal A}(x)-1] & = & -\,\frac{3}{16\pi^2}\int_0^\infty \!\!dy\, 
y\, 
{\cal K}_{\cal A}(x,y)\, \sigma_{\cal V}(y),\\ 
&& \label{AEDSE}\\ 
\nonumber {\cal B}(x) & = & - \,\frac{3}{16\pi^2} \int_0^\infty \!\!dy\, 
y\, {\cal K}_{\cal B}(x,y)\, \sigma_{\cal S}(y),\\ 
&& \label{BEDSE} 
\end{eqnarray} 
and solved numerically by iteration. 
 
To illustrate the accuracy attainable with this procedure we: evaluated the 
integrals in Eqs.~(\ref{KAang}), (\ref{KBang}) numerically for spacelike $P^2$; 
inserted ${\cal A}(k^2)\equiv 1$ and ${\cal B}(k^2)=M$ on the r.h.s.\ of 
Eqs.~(\ref{AEDSE}) and (\ref{BEDSE}); and calculated the integral over $y$ 
numerically.  This yields the estimate of the one-loop-corrected nucleon 
propagator in the spacelike region depicted in Figs.~\ref{cfA}, \ref{cfB}. The 
agreement with the algebraic result is exact. 
 
The self-consistent solution of Eqs.~(\ref{AEDSE}), (\ref{BEDSE}) in the 
spacelike region is easily obtained by iteration: the one-loop corrected 
functions are inserted on the r.h.s.\ to obtain the second iterate, which is 
then inserted on the r.h.s.\ to obtain the third iterate, etc., with the 
procedure repeated until the input and output agree within a specified 
tolerance.  That happens very quickly, with the fourth iterate from free 
nucleon seed functions (${\cal A} =1$, ${\cal B}=M$) agreeing with the third 
iterate to better than $10^{-4}\,$\%.  Hence ``three pions in the air'' are 
sufficient to fully dress the nucleon.  The functions obtained in this 
self-consistent solution are also plotted in Figs.~\ref{cfA}, \ref{cfB}: only 
${\cal A}(t^2)$ is noticeably modified cf.\ the one-loop result. 
 
To locate the mass pole in the nonperturbatively dressed nucleon propagator,
Eqs.~(\ref{AEDSE}), (\ref{BEDSE}) must also be solved for timelike $P^2$.
That requires an analytic continuation of the kernels in Eqs.~(\ref{KAang}),
(\ref{KBang}).  The primary nonanalytic feature in their integrands is the
pion pole and in continuing to timelike $P^2$ it is necessary to properly
incorporate its effect.  That is difficult when the kernels are only known
numerically and an expeditious alternative is to develop an algebraic
approximation, which is the approach we adopt.
 
It is apparent that both kernels can be considered as a sum of two terms. The 
first is proportional to the angular average of $g_{PV}^2((P-k)^2)$, and using 
Eq.~(\ref{gPVexp}) that integral can be evaluated exactly: 
\begin{eqnarray} 
\nonumber 
\bar g_{PV}^2(P^2,k^2) & := & \int d\Omega_k \,g_{PV}^2((P-k)^2)\\ 
\nonumber & = &  \frac{g^2}{4 M^2}\, {\rm e}^{- 2 (P^2+k^2)/\Lambda^2} \, 
\frac{\Lambda^2}{2 P k} \, {\rm I}_1(4 P k/\Lambda^2)\,,\\ 
&& 
\end{eqnarray} 
where ${\rm I}_1(x)$ is a modified Bessel function and $P = \sqrt{P^2}$, 
$k=\sqrt{k^2}$.  The second term in both cases is proportional to 
\begin{eqnarray} 
\omega_{g^2}(P^2,k^2) & := &\int d\Omega_k  \, \frac{g_{PV}^2((P-k)^2)} 
{(P-k)^2 + m_\pi^2}\,, 
\end{eqnarray} 
which, in general, cannot be expressed as a finite sum of known functions. 
However, if $g_{PV}$ is regular at $P=k$ and its analytic structure is not a 
key influence on the solution, then the approximation 
\begin{eqnarray} 
\nonumber \omega_{g^2}(P^2,k^2) & \approx & g_{PV}^2(|P^2-k^2|) 
\int\! d\Omega_k\, \frac{1}{(P-k)^2 + m_\pi^2} \\ 
&=& g_{PV}^2(|P^2-k^2|)\,  \frac{1}{a+\sqrt{a^2-b^2}}  \label{omegaB}\\ 
& =: & \tilde g_{PV}^2(P^2,k^2) \,\frac{1}{a+\sqrt{a^2-b^2}}\,, 
\end{eqnarray} 
where $a= P^2 + k^2 +m_\pi^2$, $b= 2 P k$, is a reliable 
tool~\cite{angleapprox}.  As these preconditions are obviously satisfied in our 
application -- the dominant physical effect in $\pi N$ physics is the pion pole 
and that appears at a mass-scale much lower than those present in $g_{PV}$ -- 
we pursue our analysis using the following algebraic approximations: 
\begin{eqnarray} 
\nonumber 
\tilde {\cal K}_{\cal A}(x,y)   & = & -\, \frac{1}{2}\, \bar g_{PV}^2(x,y)\,(x+y) \\ 
\nonumber &  &  +\, \tilde g_{PV}^2(x,y)\,\frac{(x-y)^2 + m_\pi^2 
(x+y)}{a+\sqrt{a^2-b^2}}\,, \\ 
&& \label{KAapprox}\\ 
\nonumber 
\tilde {\cal K}_{\cal B}(x,y)  &= & \bar g_{PV}^2(x,y) \\ 
& & - \,\tilde g_{PV}^2(x,y)\,\frac{ 2 m_\pi^2}{a+\sqrt{a^2-b^2}} \,. 
\label{KBapprox} 
\end{eqnarray} 
To illustrate their efficacy, in Figs.~\ref{cfA}, \ref{cfB} we plot the 
self-consistent solutions of 
\begin{eqnarray} 
\nonumber x [{\cal A}(x)-1] & = & -\,\frac{3}{16\pi^2}\int_0^\infty \!\!dy\, 
y\, \tilde {\cal K}_{\cal A}(x,y)\, \sigma_{\cal V}(y),\\ 
\label{AEDSEapp} && \\ 
\nonumber {\cal B}(x) & = & - \,\frac{3}{16\pi^2} \int_0^\infty \!\!dy\, y\, 
\tilde{\cal K}_{\cal B}(x,y)\, \sigma_{\cal S}(y).\\ 
& & \label{BEDSEapp} 
\end{eqnarray} 
The error introduced by the approximation is never more than $1$\% and is only 
that large for ${\cal A}(t^2=0)$. 
 
We can now define the model's analytic continuation to the timelike region. The 
approximate kernels' primary nonanalyticity is a square-root branch point whose 
appearance and location are tied to the simple pole in the pion propagator, and 
in continuing to $P^2<0$ it is necessary to include the discontinuity across 
the associated cut.  That is accomplished~\cite{fukuda76} by adding the 
following additional terms to the r.h.s.\ of Eqs.~(\ref{AEDSEapp}), 
(\ref{BEDSEapp}), respectively: 
\begin{eqnarray} 
\label{nonanA} & & - \frac{3}{16 \pi^2}\int_{x_b}^0\!\! dy\, y\, 
\hat g^2(x,y)\,\Delta\tilde{\cal K}_{\cal A}(x,y)\, 
\sigma_{\cal V}(y) \,,\\ 
\label{nonanB}&& - \frac{3}{16 \pi^2}\int_{x_b}^0\!\! dy\, y\, 
\hat g^2(x,y)\,\Delta\tilde{\cal K}_{\cal B}(x,y)\, \sigma_{\cal S}(y)\, , 
\end{eqnarray} 
where 
\begin{equation} 
 \Delta\tilde{\cal K}_{\cal A}(x,y) = -\,\frac{\Delta{\cal K}_{\cal 
B}(x,y)}{2\,m_\pi^2}\, \left[ (x-y)^2 + m_\pi^2 (x+y) \right]\,, 
\end{equation} 
\begin{eqnarray} 
\Delta\tilde{\cal K}_{\cal B}(x,y) & = & m_\pi^2\,\frac{\sqrt{(x+y+m_\pi^2)^2- 
4 x y}}{x \,y}\,, 
\end{eqnarray} 
and $y=x_b=-(\sqrt{-x}-m_\pi)^2$ is the location of the branch point.  (NB.\ 
These terms are present only when $P^2+m_\pi^2<0$.)   The self-consistent 
solutions of Eqs.~(\ref{AEDSEapp})-(\ref{nonanB}) are depicted in 
Figs.~\ref{cfA}, \ref{cfB} and unsurprisingly there is little difference 
between the one-loop results and the self-consistent solution. 
 
In Fig.~\ref{Eshift} we compare the exact one-loop mass shift with that 
obtained numerically using the approximate kernels.  The error is never more 
than $5$\% with the approximation always overestimating the magnitude of the 
shift. (It is noteworthy that a large part of the one-loop mass-shift is due to 
the vector self energy; e.g., with $\Lambda=0.9\,$GeV, $(\delta M_+)_{\rm 
one-loop}$ is $40\,$\% smaller if the vector self-energy is neglected.) 
 
The fully dressed nucleon mass, $M_D$, is obtained by solving 
\begin{equation} 
M_D^2 {\cal A}^2(-M_D^2) = [M + {\cal B}(-M_D^2)]^2 
\end{equation} 
with the nonperturbative mass shift given by $\delta M_+ = M_D - M$. Again, 
this definition is completely equivalent to Eq.~(\ref{defndeltaMp}) evaluated 
at $M_D$ with the self-consistent solution of the DSE.  The 
$\Lambda$-dependence of the nonperturbative shift is also depicted in 
Fig.~\ref{Eshift} and comparison with the numerical one-loop result shows that 
the additional pion dressing adds $\lesssim 5\,$\% to $|\delta M_+|$. 
 
Thus far we have used our Euclidean model to quantitatively reproduce the 
perturbative results of Secs.~\ref{subsec:modelfieldtheory}, 
\ref{subsec:modelfieldtheorynonlinear} and thereby make transparent the 
equivalence of the Euclidean and Minkowski formulations.  In addition we have 
shown that the one-loop mass shift is $\sim 95$\% of the total. 
 
\begin{figure}[t] 
\centerline{\includegraphics[height=6.5cm]{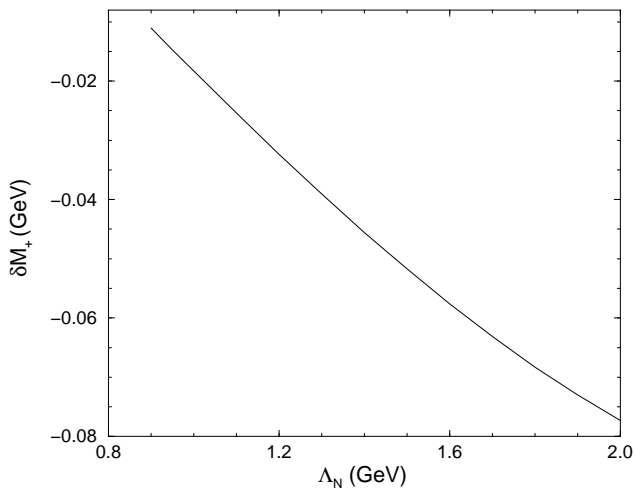}} 
\caption{\label{offshell}  Effect on the mass-shift produced by including 
nucleon-off-shell suppression in the $\pi N N$ vertex: 
Eq.~(\protect\ref{gNoff}), for $\Lambda =0.9\,$GeV, $m_\pi=0.14\,$GeV.  (NB.\ 
$\delta M_+(\Lambda=0.9\,{\rm GeV},\Lambda_N=\infty) = -0.14\,$GeV.)} 
\end{figure}

However, we have not yet considered an effect of nuc\-leon compositeness.  A 
covariant $\pi N N$ vertex function must depend on three independent variables: 
\mbox{$g_{PV}= g_{PV}(P^2,k^2,(P-k)^2)$}~\cite{mr97}, and hitherto we have 
neglected its dependence on $P^2$, $k^2$.  (We have already seen that $g_{PV} = 
g_{PV}((P-k)^2)$ corresponds to a Pauli-Villars regularisation of the pion 
propagator alone.)  The calculation of form factors that describe interactions 
between composite objects; e.g., studies of the $\rho$-$\omega$ mass 
splitting~\cite{rhopipipeter,rhopipi} and electromagnetic form 
factors~\cite{pieterGamma,revbasti,weiss,gstarpig}, indicates that the $\pi N 
N$ vertex should also suppress the pion-nucleon coupling when the nucleons are 
off-shell. We conduct an initial, exploratory study of this effect by 
considering the product {\it Ansatz} 
\begin{eqnarray} 
\nonumber\lefteqn{g_{PV}(P^2,k^2,P\cdot k) =  } \\ 
& & \frac{g}{2 M} \, {\rm e}^{-(P-k)^2/\Lambda^2}\, {\rm 
e}^{-(P^2+M^2+k^2+M^2)/\Lambda_N^2}\,, \label{gNoff} 
\end{eqnarray} 
which reduces to Eq.~(\ref{gPVexp}) when $\Lambda_N\to \infty$ and guarantees 
$g_{PV}(-M^2,-M^2,0)= g/(2 M)$, as required.  Previous applications of such a 
form factor in the $\pi N$ sector~\cite{pearcea} typically require 
\begin{equation} 
\Lambda_N /\Lambda\sim 1.5\,-2.0\,. 
\end{equation} 
(NB.\ While $\Lambda_N$ is calculable using a covariant model of the nucleon, 
no such calculations exist and to constrain its value we must currently rely on 
phenomenology.)  The effect on the mass shift of this off-shell suppression is 
depicted in Fig.~\ref{offshell}: it is significant, leading to a reduction of 
$\gtrsim 50\,$\% in $|\delta M_+|$. For $\Lambda_N \to \infty$; i.e., in the 
absence of the off-shell suppression, this effect can be mimicked by a 
reduction in $\Lambda$; e.g., $\Lambda \to \Lambda^\prime=0.7\,$GeV yields 
$\delta M_+=-0.07\,$GeV, and we note that $\Lambda^\prime/\sqrt{2} = 0.5\,$GeV, 
which is commensurate with $\lambda_{\rm CBM}\approx 0.4\,$GeV, after 
Eq.~(\ref{CBMshift}). 
 
Combining all the elements of our analysis we arrive at a result for the shift 
in the nucleon's mass owing to the $\pi N$-loop (for $g_A=1.26$, in GeV): 
\begin{equation} 
\label{shiftF} - \delta M_+ \simeq  ( 0.039 - 0.063 ) \, g_A^2 = (0.061 - 
0.099)\,. 
\end{equation} 
 
In the preceding, for illustrative clarity, we did not account for the
effects of finite vertex renormalisation; i.e., we set $Z_1=1=Z_2$ in
Eq.~(\ref{EDSE}).  Studies using the CBM indicate that a quantitative
description of $\pi N$ vertex renormalisation requires that the $\Delta$ be
treated on an equal footing with the nucleon and that this is crucial to
obtaining a convergent expansion~\cite{tonyCBM,Dodd:pw}.  Indeed, one finds,
as here, that the $\pi$-loop acts to suppress the nucleon's wave-function
renormalisation; i.e., it forces $Z_2 < 1$, but in the CBM this effect is
compensated by an almost matching suppression of $Z_1$ so that the bare and
renormalised $\pi N$ couplings are little different.  A self-consistent,
covariant treatment of the coupled composite-$N$-$\Delta$ system is more than
we are able to describe herein.  However, the CBM studies suggest that a
reliable estimate of the effect of including the $\Delta$ can be obtained
simply by solving an analogue of Eq.~(\ref{EDSE}) with $Z_1=Z_2$ for a
renormalised model.

We have done this and thereby arrive at a robust result: the $\pi N$-loop
reduces the nucleon's mass by $\sim 10\,$-$\,20$\%~\cite{fn:error}.  Extant
calculations; e.g., Refs.~\cite{tonyANU,tonyCBM,harry}, show that the
contribution from the analogous $\pi \Delta$-loop is of the same sign and no
greater in magnitude so that the likely total reduction is $20\,$-$\,40$\%.
Based on these same calculations we anticipate that the $\Delta$ mass is also
reduced by $\pi$ loops but by a smaller amount ($\sim 50\,$-$\,100\,$MeV
less).
 
How does that affect the quark-diquark picture of baryons? To address this 
issue we again solved the Faddeev equations, this time requiring that the 
quark-diquark component yield higher masses for the $N$ and $\Delta$: $M_N=0.94 
+ 0.2=1.14\,$GeV, $M_\Delta= 1.232+0.1=1.332\,$GeV. The results, presented in 
the third and fourth rows of Table~\ref{tableMass}, establish that the effects 
are not large.  In this case omitting the axial-vector diquark yields 
$M_N=1.44\,$GeV, which signals a $10\,$\% increase in the importance of the 
scalar-diquark component of the nucleon. (It is an \textit{increase} because 
this component now requires less correction.  Note, too, that the scalar 
diquark's charge radius, $r_{0^+}=0.63\,$fm, is $15\,$\% larger.)  It also 
announces a reduction in the role played by axial-vector diquark correlations 
in the nucleon, since now restoring them only reduces the nucleon's core mass 
by $21$\%, with $\pi$ self-energy corrections providing the remaining $14$\%. 
It is thus apparent that requiring an exact fit to the $N$ and $\Delta$ masses 
using only quark and diquark degrees of freedom leads to an overestimate of the 
role played by axial-vector diquark correlations: it forces the $1^+$ diquark 
to mimic, in part, the effect of pions since they both act to reduce the mass 
cf.\ that of a quark$+$scalar-diquark baryon. 
 
\section{Epilogue} 
\label{sec:epil} 
We showed that an internally consistent description of the $N$ and $\Delta$ 
masses is easily obtained using a Poincar\'e covariant Faddeev equation that 
represents baryons as composites of a confined-quark and -diquark.  We term 
this the ``core mass'' of the baryons.  They are weakly bound in the limited 
sense that the sum of the masses of their primary constituents is little 
greater than their core mass. 
 
The on-shell $\pi N N$ and $\pi N \Delta$ couplings are large and hence it is
conceivable that $\pi N$ and $\pi \Delta$ self-energy corrections to the
nucleon's mass may be significant.  We therefore studied the effects of the
$\pi N$-loop on the nucleon's core mass and found that, in well-constrained
models, this loop reduces that mass by $\lesssim 20$\%.  Including the $\pi
\Delta$ self-energy contribution, the total reduction is likely to be between
$20$ and $40$\%. While this is a material effect it does not undermine the
qualitative picture of baryons suggested by the Faddeev equation; namely,
that baryons are primarily quark-diquark composites.  This is consistent with
the fact that a converged nonperturbative calculation of the $\pi$-induced
self-energy requires only three ``pions in the air,'' but to be certain we
re-solved the Faddeev equation aiming at nucleon and $\Delta$ masses
corrected for the $\pi$ self-energy contribution, and found little change in
the character of the solution.
 
The one notable effect was a material reduction in the nucleon's axial-vector 
diquark component.  This is easily understood: ignoring $\pi$-loops forces the 
axial-vector diquarks to mimic their effect.  That surrogacy cannot be 
completely effective and may have led to quantitative errors, and errors of 
interpretation, in contemporary quark-diquark based calculations of quantities 
such as the neutron's charge form factor and the ratio $\mu_p G_E^p/G_M^p$. Our 
results should serve as a signal of this possibility and stimulate increased 
caution and an objective reanalysis. 
 
Our exploration of the role of $\pi$-loops was pedagogical.  We made clear that 
the leading nonanalytic contribution to the nucleon's mass arises from that 
part of the loop integral which corresponds to a positive-energy nucleon; i.e., 
whether the $\pi N N$ coupling is pseudoscalar or pseudovector, the 
$Z$-diagrams do not affect the leading nonanalytic behaviour. Furthermore, we 
showed explicitly that the one-loop mass shift calculated with a pseudoscalar 
coupling is {\it precisely} the same as that obtained with a pseudovector 
coupling, so long as, and only if, no diagrams are overlooked in the 
pseudoscalar calculation.  We illustrated that, using any translationally 
invariant regularisation procedure which preserves information about the pion's 
finite size, the tadpole (Hartree) diagram generated by a pseudoscalar coupling 
cannot be neglected because it balances the very large contribution from the 
pseudoscalar $Z$-diagram.  This result should not be overlooked in the 
phenomenological application of model field theories founded on hadronic 
degrees of freedom. 
 
\begin{acknowledgments} 
CDR is grateful for the hospitality of the staff at the UK Institute for 
Particle Physics Phenomenology, University of Durham, during a visit in which 
part of this work was conducted and for financial support from the Institute, 
and for the generosity and warmth of the Members of the Senior Common Room, 
Grey College, Durham; CDR and PCT are grateful for the hospitality of the staff 
at the Special Research Centre for the Subatomic Structure of Matter during 
visits in which part of this research was conducted and also for financial 
support provided by the Centre; MO, CDR and PCT are grateful for financial 
support from the Europ\"aisches Graduiertenkolleg: Basel-T\"ubingen, ``Hadronen 
im Vakuum, in Kernen und Sternen'' and Universit\"at T\"ubingen; and M.~Oettel 
is grateful for financial support from the A.~v.~Humboldt foundation. This work 
was supported by: the US Department of Energy, Nuclear Physics Division, under 
contract no.~\mbox{W-31-109-ENG-38}; the Deutsche Forschungsgemeinschaft, under 
contract no.\ SCHM~1342/3-1; the US National Science Foundation, under grant 
nos.\ PHY-0071361 and PHY-9722429; the Australian Research Council; and 
Adelaide University; and benefited from the resources of the US National Energy 
Research Scientific Computing Center. 
\end{acknowledgments} 
 
\appendix 
\section{Euclidean Conventions} 
\label{App:EM} 
\subsection{Metric and Spinors} 
In our Euclidean formulation: 
\begin{equation} 
p\cdot q=\sum_{i=1}^4 p_i q_i\,; 
\end{equation} 
\begin{eqnarray}
& \{\gamma_\mu,\gamma_\nu\}=2\,\delta_{\mu\nu}\,;\; 
\gamma_\mu^\dagger = \gamma_\mu\,;\; 
\sigma_{\mu\nu}= \sfrac{i}{2}[\gamma_\mu,\gamma_\nu]\,; &  \\
& {\rm tr}_[\gamma_5\gamma_\mu\gamma_\nu\gamma_\rho\gamma_\sigma]= 
-4\,\epsilon_{\mu\nu\rho\sigma}\,, \epsilon_{1234}= 1\,. & 
\end{eqnarray}

A positive energy spinor satisfies 
\begin{equation} 
\bar u(P,s)\, (i \gamma\cdot P + M) = 0 = (i\gamma\cdot P + M)\, u(P,s)\,, 
\end{equation} 
where $s=\pm$ is the spin label.  It is normalised: 
\begin{equation} 
\bar u(P,s) \, u(P,s) = 2 M 
\end{equation} 
and may be expressed explicitly: 
\begin{equation} 
u(P,s) = \sqrt{M- i {\cal E}}\left( 
\begin{array}{l} 
\chi_s\\ 
\displaystyle \frac{\vec{\sigma}\cdot \vec{P}}{M - i {\cal E}} \chi_s 
\end{array} 
\right)\,, 
\end{equation} 
with ${\cal E} = i \sqrt{\vec{P}^2 + M^2}$, 
\begin{equation} 
\chi_+ = \left( \begin{array}{c} 1 \\ 0  \end{array}\right)\,,\; 
\chi_- = \left( \begin{array}{c} 0\\ 1  \end{array}\right)\,. 
\end{equation} 
For the free-particle spinor, $\bar u(P,s)= u(P,s)^\dagger \gamma_4$. 
 
The spinor can be used to construct a positive energy projection operator: 
\begin{equation} 
\label{Lplus} \Lambda_+(P):= \frac{1}{2 M}\,\sum_{s=\pm} \, u(P,s) \, \bar 
u(P,s) = \frac{1}{2M} \left( -i \gamma\cdot P + M\right). 
\end{equation} 
 
A negative energy spinor satisfies 
\begin{equation} 
\bar v(P,s)\,(i\gamma\cdot P - M) = 0 = (i\gamma\cdot P - M) \, v(P,s)\,, 
\end{equation} 
and possesses properties and satisfies constraints obtained via obvious analogy 
with $u(P,s)$. 
 
A charge-conjugated Bethe-Salpeter amplitude is obtained via 
\begin{equation} 
\bar\Gamma(k;P) = C^\dagger \, \Gamma(-k;P)^{\rm T}\,C\,, 
\end{equation} 
where ``T'' denotes a transposing of all matrix indices and 
$C=\gamma_2\gamma_4$ is the charge conjugation matrix, $C^\dagger=-C$. 
 
In describing the $\Delta$ resonance we employ a Rarita-Schwinger spinor to 
unambiguously represent a covariant spin-$3/2$ field.  The positive energy 
spinor is defined by the following equations: 
\begin{eqnarray} 
(i \gamma\cdot P + M)\, u_\mu(P;r) & = & 0\,,\\ 
\gamma_\mu u_\mu(P;r) & = & 0\,,\\ 
P_\mu u_\mu(P;r) & = & 0\,, 
\end{eqnarray} 
where $r=-3/2,-1/2,1/2,3/2$.  It is normalised: 
\begin{equation} 
\bar u_{\mu}(P;r^\prime) \, u_\mu(P;r) = 2 M\,, 
\end{equation} 
and satisfies a completeness relation 
\begin{equation} 
\frac{1}{2 M}\sum_{r=-3/2}^{3/2} u_\mu(P;r)\,\bar u_\nu(P;r) = 
\Lambda_+(P)\,R_{\mu\nu}\,, 
\end{equation} 
where 
\begin{equation} 
R_{\mu\nu} = \delta_{\mu\nu} I_{\rm D} -\frac{1}{3} \gamma_\mu \gamma_\nu + 
\frac{2}{3} \hat P_\mu \hat P_\nu I_{\rm D} - i\frac{1}{3} [ \hat P_\mu 
\gamma_\nu - \hat P_\nu \gamma_\mu]\,, 
\end{equation} 
with $\hat P^2 = -1$, which is very useful in simplifying the positive energy 
$\Delta$'s Faddeev equation. 
 
\subsection{Euclidean One-loop Calculations} 
In Eqs.~(\ref{AEoneloop}) and (\ref{BEoneloop}) 
\begin{eqnarray} 
a(t,\kappa) & = & - a_0(t,\kappa) + a_1(t,\kappa) + a_2(t,\kappa)\,,\\ 
b(t,\kappa) & = & a_0(t,\kappa) + a_2(t,\kappa)\,, 
\end{eqnarray} 
where 
\begin{eqnarray} 
\nonumber 
\lefteqn{ a_0(t,\kappa) =  }\\ 
& &  \kappa^2 \bigg[ {\cal I}_0(\omega_\pi(\kappa)) \, 
\nonumber \sum_{s=0}^1 \frac{\omega_N(\kappa)+ (-)^s \omega_\pi(\kappa)} 
{(\omega_N(\kappa)+ (-)^s \omega_\pi(\kappa))^2 + t^2} \\ 
\nonumber & & - {\cal I}_0(\Omega_N(t,\kappa)) \, 
\frac{\omega_\pi(\kappa)}{\Omega_N^2(t,\kappa) -\omega_\pi^2(\kappa)}\\ 
 & & - {\cal I}_0(\bar\Omega_N(t,\kappa)) \, 
\frac{\omega_\pi(\kappa)}{\bar\Omega_N^2(t,\kappa) -\omega_\pi^2(\kappa)} 
\bigg] , \label{a0tk} 
\end{eqnarray} 
\begin{eqnarray} 
\nonumber 
\lefteqn{ a_1(t,\kappa) = m_\pi^2\, \omega_\pi(\kappa)}\\ 
&\times & \nonumber \bigg[ {\cal I}_0(\omega_\pi(\kappa)) \, \sum_{s=0}^1 
\frac{(-)^s} 
{(\omega_N(\kappa)+ (-)^s \omega_\pi(\kappa))^2 + t^2} \\ 
\nonumber && + \frac{i}{t}\, {\cal I}_0(\Omega_N(t,\kappa)) \, 
\frac{\Omega_N(t,\kappa)}{\Omega_N^2(t,\kappa) -\omega_\pi^2(\kappa)}\\ 
\nonumber & & - \frac{i}{t}\, {\cal I}_0(\bar\Omega_N(t,\kappa)) \, 
\frac{\bar\Omega_N(t,\kappa)}{\bar\Omega_N^2(t,\kappa) -\omega_\pi^2(\kappa)} 
\bigg] ,\\ 
\nonumber & -  &  \frac{i}{t}\, \omega_\pi(\kappa)\, \bigg[ 
\Omega_N(t,\kappa) \,{\cal I}_0(\Omega_N(t,\kappa)) \\ 
 && - \bar\Omega_N(t,\kappa) \,{\cal I}_0(\bar\Omega_N(t,\kappa)) 
\bigg] 
\end{eqnarray} 
\begin{eqnarray} 
\nonumber 
\lefteqn{ a_2(t,\kappa) = }\\ 
&   & \bigg[ {\cal I}_2(\omega_\pi(\kappa)) \, 
\nonumber \sum_{s=0}^1 \frac{\omega_N(\kappa)+ (-)^s \omega_\pi(\kappa)} 
{(\omega_N(\kappa)+ (-)^s \omega_\pi(\kappa))^2 + t^2} \\ 
\nonumber & & - {\cal I}_2(\Omega_N(t,\kappa)) \, 
\frac{\omega_\pi(\kappa)}{\Omega_N^2(t,\kappa) -\omega_\pi^2(\kappa)}\\ 
 & & - {\cal I}_2(\bar\Omega_N(t,\kappa)) \, 
\frac{\omega_\pi(\kappa)}{\bar\Omega_N^2(t,\kappa) -\omega_\pi^2(\kappa)} 
\bigg] , \label{a2tk} 
\end{eqnarray} 
where $\Omega_N(t,\kappa) = \omega_N(\kappa) + i t$, $\bar\Omega_N(t,\kappa) = 
\omega_N(\kappa) - i t$, and 
\begin{eqnarray} 
\nonumber {\cal I}_0(x) & := & 
 2 \int_0^\infty du \, {\rm e}^{-2 u^2/\Lambda^2} \frac{1}{x - i u}\\ 
\label{calI0} & = & {\rm e}^{\,2 x^2/\Lambda^2} \, 
{\rm erfc}(x\sqrt{2}/\Lambda)\,,\\ 
\nonumber {\cal I}_2(x) & := & 
2 \int_0^\infty du \, {\rm e}^{-2 u^2/\Lambda^2} \frac{u^2}{x - i u}\\ 
\label{calI2} & = & x \bigg (\frac{\Lambda}{\sqrt{2 \pi}} - x \,{\cal I}_0(x) 
\bigg)\,, 
\end{eqnarray} 
where ${\rm erfc}(x)$ is the complementary error function and both these 
functions are odd under $x\to -x$. 
 

\begin{thebibliography}{99}
\bibitem{fn:Euclidean} Herein we almost exclusively employ a Euclidean 
metric, which is described in Appendix~\ref{App:EM}.  A Minkowski metric is 
used only in sections~\ref{subsec:modelfieldtheory}, 
\ref{subsec:modelfieldtheorynonlinear}.
\bibitem{fn:confinement} This is a sufficient condition for confinement 
because of the associated violation of reflection positivity, as discussed, 
e.g., 
%
in Sec.~6.2 of Ref.~\cite{cdragw}, Sec.~2.2 of Ref.~\cite{revbasti} and 
Sec.~2.4 of Ref.~\cite{revreinhard}.
\bibitem{regwrong} The sign before the integral in Eq.~(65) of 
Ref.~\protect\cite{gunnergcm} is incorrect.
\bibitem{fn:error} The estimate of $\delta M_+$ in 
Ref.~\protect\cite{cdrqciv} is incorrect because that calculation overlooked 
the tadpole (Hartree) term and the contribution at timelike momenta 
associated with the branch cut in the kernel.
\end{thebibliography}

\end{document}